\documentclass[12pt]{article}
\pdfoutput=1

\usepackage[margin=1.05in]{geometry}
\usepackage{amsmath,amssymb}
\usepackage{mathtools}
\usepackage{graphicx}
\usepackage{microtype}
\usepackage{placeins}
\usepackage{float}
\usepackage{tikz}
\usetikzlibrary{arrows.meta,positioning,decorations.markings}
\usetikzlibrary{snakes}
\usetikzlibrary{shapes.misc}
\usepackage[nosort]{cite}
\usepackage{todonotes}
\usepackage{framed}
\usepackage{slashed}
\usepackage[colorlinks=true,linkcolor=blue,citecolor=blue,urlcolor=blue]{hyperref}

\numberwithin{equation}{section}

\newcommand{\ii}{\mathrm{i}}
\newcommand{\dd}{\mathrm{d}}
\newcommand{\NCM}{\mathrm{NCM}}

\newcommand{\eps}{\varepsilon}
\newcommand{\bpar}{\mathfrak{b}}
\newcommand{\cC}{\mathcal{C}}
\newcommand{\cF}{\mathcal{F}}
\newcommand{\cI}{\mathcal{I}}
\newcommand{\Li}{\mathrm{Li}}
\newcommand{\Res}{\operatorname{Res}}

\begin{document}

\hypersetup{pageanchor=false}

\begin{titlepage}
\thispagestyle{empty}

\vspace*{-1.3cm}

% Optional preprint number:
% \noindent\hfill BIMSA-TH-26-XX

\vspace{1.4cm}

\begin{center}

{\Large\bfseries
Non-critical M-Theory and the Resolved Conifold
\par}

\vspace{0.15cm}

{\large\bfseries
Refinement and Nonperturbative Completion
\par}
\vspace{0.9cm}
{\large Fengjun Xu\par}

\vspace{0.35cm}

{\small
Beijing Institute of Mathematical Sciences and Applications (BIMSA),\\
Beijing 101408, China
\par}

\vspace{0.22cm}

{\small
\texttt{fengjunxu@bimsa.cn}
\par}

\end{center}

\vfill

\begin{center}
\begin{minipage}{0.88\textwidth}

\begin{abstract}
We extend the Ho\v{r}ava--Keeler correspondence between finite-temperature
non-critical M-theory and the resolved-conifold A-model to the refined theory.
The grand potential of the stationary rotating non-critical M-theory vacuum reproduces the
refined Gopakumar--Vafa expansion, with the angular chemical potential
deforming the $\Omega$-background away from the self-dual locus
$\eps_1=-\eps_2$.  At finite thermal radius, we establish an exact match
with the nonperturbative conifold completions proposed by Hattab--Palti and
by Chuang in the unrefined and refined cases, respectively.  Both the
one-particle Schwinger integrand and its integration cycle arise from the
non-critical M-theory spectrum and resolvent, providing a microscopic spectral realization
of these completions.  The correspondence fixes the nonconstant BPS sector
and the cubic local contribution, while the remaining polynomial ambiguity
and the modulus-independent constant-map sector require separate
normalization.  We explain why the single primitive spinless multiplet of
the resolved conifold makes this identification possible, and discuss what
additional charge- and spin-dependent microscopic data would be needed for
an extension to more general local Calabi--Yau geometries.
\end{abstract}

\end{minipage}
\end{center}

\vfill

\end{titlepage}

\hypersetup{pageanchor=true}

\tableofcontents
\clearpage

% ===========================================================================

\section{Introduction}
\label{sec:intro}

Non-critical M-theory (NCM), introduced by Ho\v{r}ava and Keeler, is the
double-scaled theory of noninteracting fermions moving in an isotropic inverted
harmonic potential on a two-dimensional eigenvalue plane
\cite{Horava:2005tt, Horava:2005wm,Horava:2007ds}.  Its vacua are distinguished by how the Fermi sea is
filled.  The familiar two-dimensional string theories arise from special
fillings: restricting to a fixed angular-momentum sector $J=q$ gives the
type~0A string at RR flux $q$; fixing the energy and parity of one Cartesian
oscillator gives type~0B, while a one-sided filling of the remaining
oscillator gives the bosonic $c=1$ string.  By contrast, filling every
angular-momentum sector to a common depth $\mu$ defines the all-$q$
M-vacuum, a genuinely $2{+}1$-dimensional state rather than a
two-dimensional string background.  We study this all-sector filling and
its stationary rotating deformation.

Non-critical string theories are among the rare examples in which genuinely stringy dynamics can be solved exactly, largely because they admit dual descriptions in terms of matrix quantum mechanics/matrix models.  The subject has also been revived in recent years, ranging from the nonperturbative
completion of the Liouville and $c=1$ strings by worldsheet D-instantons
\cite{Balthazar:2019rnh,Sen:2019qqg,Eniceicu:2022nay} to newly solvable
two-dimensional models such as the Virasoro minimal string \cite{Collier:2023cyw} and complex Liouville strings \cite{Collier:2024kmo}. Viewed from the standpoint of critical-string theory, however, a non-critical string
theory usually does not by itself specify a complete background\footnote{We thank C. Vafa for emphasizing this perspective on non-critical
strings. Nevertheless, it remains an open question whether quantum-gravity consistency conditions familiar from the swampland program have an analogue in non-critical strings, perhaps through their Liouville/linear-dilaton sector.}; rather, it isolates a universal decoupled sector associated with a singular double-scaling limit. Relations to topological strings provide a concrete illustration: an exactly solvable matrix model can describe a non-critical string while the same amplitudes simultaneously encode a protected sector of a higher-dimensional string compactification.  The canonical example is the Ghoshal--Vafa correspondence \cite{Ghoshal:1995wm}, in which the self-dual $c=1$ string reproduces the conifold sector of the topological B-model.  From this perspective, the solvable non-critical theory captures a distinguished sector of a larger string construction rather than specifying that construction in its entirety.  A closely related deformed-conifold B-model description has been proposed for the self-dual super-affine type~0B theory \cite{Ita:2004topological}.  Since both the $c=1$ and type~0B theories arise from special fillings of non-critical M-theory, these examples suggest a vacuum-dependent picture in which different fermionic fillings isolate different protected sectors of a broader string framework.

Indeed, the all-$q$ M-vacuum realizes an A-model counterpart of the correspondence proposed by Ho\v{r}ava and Keeler \cite{Horava:2005wm}: after compactifying Euclidean time on a
circle of radius $R$, they argued that the grand potential of their vacuum has the same weak-coupling expansion as the nonconstant BPS sector of the closed topological A-model on the resolved conifold, once local and modulus-independent
normalization data are separated. Keeping the oscillator scale
$\omega_0$ explicit, the positive Schwinger coupling and complexified
K\"ahler modulus are
\begin{equation}
 \lambda=2\pi R\omega_0,
 \qquad
 t=2\pi R\mu+\ii\pi .
 \label{eq:intro-HK-map}
\end{equation}
Thus the thermal radius $R$ sets the coupling in oscillator units, while the
Fermi depth $\mu$ determines the K\"ahler modulus, whose imaginary part we fix once and for all to the
antiperiodic value $\ii\pi$.\footnote{More generally, NCM allows
$t=2\pi R\mu+\ii\hat\vartheta$ with $\hat\vartheta\sim\hat\vartheta+2\pi$; the choice
relevant here is $\hat\vartheta=\pi$. Ho\v{r}ava and Keeler use the equivalent
coupling $g_A=\ii\lambda$.}

Because this vacuum retains every angular-momentum sector, the
correspondence can also be probed by changing their relative occupations.
An angular chemical potential $\Omega$ does precisely this while
preserving stationarity, giving the rotating family of Ho\v{r}ava and
Keeler \cite{Horava:2005tt}.  With $h$ the one-particle Hamiltonian, the
Fermi surface becomes
\begin{equation}
 h+\Omega J=-\mu .
 \label{eq:intro-rotating-vacuum}
\end{equation}
The resulting spectral kernel contains two split oscillator weights,
\begin{equation}
 \alpha_\pm=\omega_0\pm\ii\Omega .
 \label{eq:intro-alpha}
\end{equation}
This splitting is precisely the structure expected from refinement. The refined topological string replaces the self-dual graviphoton background of the ordinary theory by a two-parameter equivariant background, characterized by $\eps_1$ and $\eps_2$, with the unrefined theory recovered on the self-dual locus $\eps_1=-\eps_2$ \cite{Awata:2005fa,Iqbal:2007ii,Awata:2008ed,Aganagic:2012hs}. Multiplying the two NCM frequencies by the circumference of the thermal circle then gives the natural identification
\begin{equation}
\eps_1=2\pi R\alpha_+,
\qquad
-\eps_2=2\pi R\alpha_- ,
\label{eq}
\end{equation}
with the same K\"ahler modulus $t$ as in \eqref{eq}. We will show that, under this identification, the rotating grand potential reproduces the refined Gopakumar--Vafa expansion. The refinement parameters therefore acquire a direct spectral interpretation as the two frequencies resolved by the rotating NCM vacuum.

The deeper question is whether these correspondences extend beyond their
weak-coupling expansions.  The NCM grand potential is defined at finite
$R$, before any expansion is performed.  This led Ho\v{r}ava and Keeler to
conjecture that NCM provides a nonperturbative definition of the
resolved-conifold A-model \cite{Horava:2005wm}.  A precise target for
testing this conjecture is supplied by the new Schwinger-contour construction
of Hattab and Palti \cite{Hattab:2024chf,Hattab:2024ewk}.  In the
Gopakumar--Vafa description, the nonconstant A-model free energy is the
one-loop Schwinger determinant of wrapped M2-brane BPS states
\cite{Gopakumar:1998ii,Gopakumar:1998jq}.  The usual perturbative
Gopakumar--Vafa expansion is obtained from one family of poles of this
Schwinger representation, giving the familiar multicover series.
Hattab and Palti instead retain the oriented contour in complexified
proper time, thereby incorporating the additional pole contributions
that are exponentially small and invisible to the genus expansion
\cite{Hattab:2024chf,Hattab:2024ewk}.  Chuang extends this construction to
the refined theory with two equivariant weights \cite{Chuang:2025aaa}.

We show that the NCM spectral problem supplies both the Schwinger integrand and the integration cycle required for these completions.  A symmetric Wick rotation of the real proper-time representation of the NCM grand potential gives precisely the Schwinger representation of the resolved-conifold free energy, with its nonzero-pole contour inherited from the original resolvent prescription.  For the unrotated vacuum this reproduces the Hattab--Palti completion \cite{Hattab:2024chf,Hattab:2024ewk}; for the rotating vacuum the same construction gives Chuang's refined completion \cite{Chuang:2025aaa}, reducing smoothly to the former as $\Omega\to0$.  The integer poles reproduce the usual Gopakumar--Vafa multicover expansion, while the additional oscillator poles generate the exponentially suppressed contributions that are invisible in the genus expansion.  In this sense, the Ho\v{r}ava--Keeler correspondence is promoted from an agreement of weak-coupling expansions to an equality of the nonperturbatively completed nonconstant resolved-conifold free energies at finite $R$, up to the local and modulus-independent normalization terms discussed below.

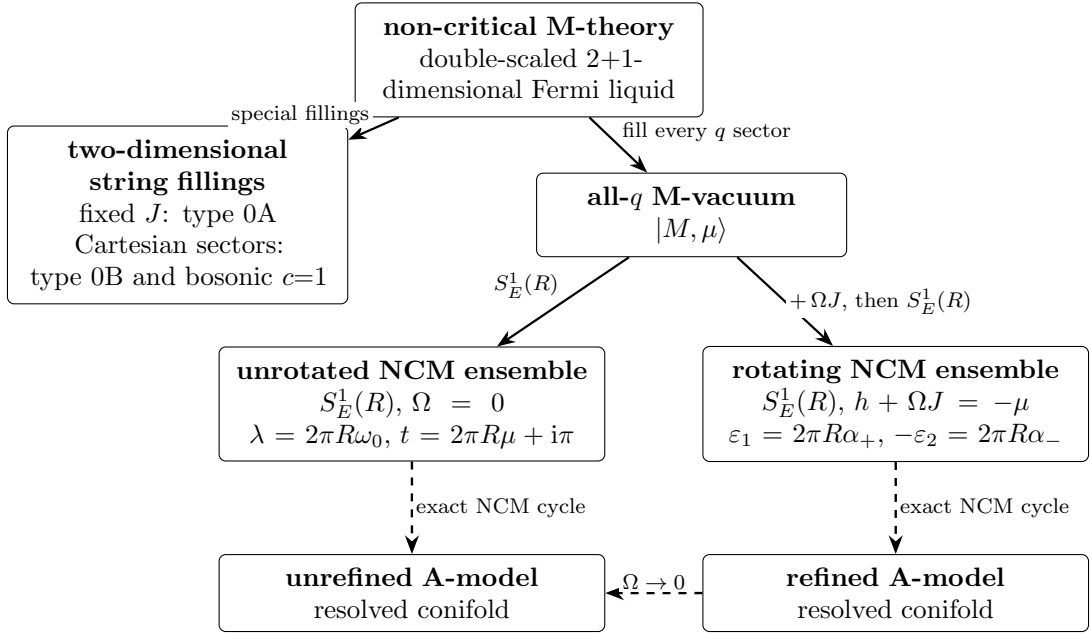
\begin{figure}[H]
\centering
\begin{tikzpicture}[>=Stealth,
  base/.style={draw, rounded corners=2.5pt, align=center, inner sep=5pt,
               font=\footnotesize, minimum height=0.95cm},
  source/.style={base, text width=4.25cm},
  side/.style={base, text width=4.15cm},
  state/.style={base, text width=3.75cm},
  thermal/.style={base, text width=4.75cm},
  amodel/.style={base, text width=4.75cm},
  lab/.style={font=\scriptsize, align=center, fill=white, inner sep=1.2pt},
  ncmarrow/.style={->, line width=0.85pt},
  corr/.style={->, dashed, line width=0.9pt}]

  \node[source] (ncm) at (0,0)
    {\textbf{non-critical M-theory}\\
     double-scaled $2{+}1$-dimensional Fermi liquid};

  \node[side] (strings) at (-4.65,-2.05)
    {\textbf{two-dimensional string fillings}\\
     fixed $J$: type~0A\\
     Cartesian sectors: type~0B and bosonic $c{=}1$};

  \node[state] (mvac) at (2.15,-2.05)
    {\textbf{all-$q$ M-vacuum}\\$|M,\mu\rangle$};

  \node[thermal] (unrot) at (-1.55,-4.55)
    {\textbf{unrotated NCM ensemble}\\
     $S_E^1(R)$, $\Omega=0$\\
     $\lambda=2\pi R\omega_0$, $t=2\pi R\mu+\ii\pi$};

  \node[thermal] (rot) at (4.85,-4.55)
    {\textbf{rotating NCM ensemble}\\
     $S_E^1(R)$, $h+\Omega J=-\mu$\\
     $\eps_1=2\pi R\alpha_+$, $-\eps_2=2\pi R\alpha_-$};

  \node[amodel] (ordinary) at (-1.55,-7.05)
    {\textbf{unrefined A-model}\\
     resolved conifold};

  \node[amodel] (refined) at (4.85,-7.05)
    {\textbf{refined A-model}\\
     resolved conifold};

  \draw[ncmarrow] (ncm) --
    node[lab, above left]{special fillings} (strings);
  \draw[ncmarrow] (ncm) --
    node[lab, above right]{fill every $q$ sector} (mvac);
  \draw[ncmarrow] (mvac) --
    node[lab, above left]{$S_E^1(R)$} (unrot);
  \draw[ncmarrow] (mvac) --
    node[lab, right]{$+\,\Omega J$, then $S_E^1(R)$} (rot);
  \draw[corr] (unrot) --
    node[lab, right]{exact NCM cycle} (ordinary);
  \draw[corr] (rot) --
    node[lab, right]{exact NCM cycle} (refined);
  \draw[corr] (refined.west) --
    node[lab, above]{$\Omega\to0$} (ordinary.east);
\end{tikzpicture}
\caption{Vacuum-dependent pattern of correspondences.  Special fillings of
NCM reproduce the two-dimensional type~0A, type~0B, and bosonic $c=1$
strings, while filling all angular-momentum sectors gives the M-vacuum
relevant here.  After thermal compactification, the unrotated and rotating
ensembles correspond, respectively, to the unrefined and refined
resolved-conifold BPS sectors.  Their exact NCM proper-time cycles give the
Hattab--Palti and Chuang contours.  Dashed arrows denote equality of the
protected Schwinger data and integration cycles, rather than an identity of
microscopic constituents or spacetime geometries.}
\label{fig:vacuum-map}
\end{figure}

The origin of this agreement can already be seen at the one-particle level.  The nonconstant resolved-conifold contribution is generated by a single primitive spin-zero M2-brane wrapped on the rigid $\mathbb P^1$, together with its orbital excitations and Kaluza--Klein tower.  On the NCM side, summing over all angular-momentum sectors of the one-particle spectrum produces exactly the same two-frequency orbital factor, while compactification of Euclidean time supplies the corresponding Kaluza--Klein sum.  Moreover, the NCM resolvent fixes the proper-time integration cycle rather than merely reproducing the perturbative integrand.  Thus the same microscopic spectral problem determines both the one-particle contribution and the prescription for its nonperturbative continuation.  This is the sense in which NCM provides a microscopic \emph{spectral} origin for the resolved-conifold amplitude; it does not imply an identification of individual NCM fermions with wrapped M2-branes.

These results sharpen the vacuum-dependent picture summarized in Figure~\ref{fig:vacuum-map}.  Special fillings of the same underlying fermionic theory give the familiar $c=1$ and type~0 string backgrounds and their associated topological-string relations, whereas the all-$q$ M-vacuum leads to the resolved-conifold A-model.  Turning on the angular chemical potential then refines this correspondence.  The choice of fermionic vacuum therefore appears to select which protected topological-string sector is realized, while the underlying NCM spectral prescription determines its nonperturbative continuation.

This note is organized as follows.  Section~\ref{sec:ncm} reviews the NCM
fillings, the finite-temperature ensemble, and the original unrefined
Ho\v{r}ava--Keeler correspondence with topological A-model.  Section~\ref{sec:refinement} introduces
the rotating vacuum, and establishes the perturbative correspondence
with the refined topological string.  Section~\ref{sec:exact-completion}
derives the exact proper-time contour and its unrefined and refined
specializations, including the treatment of the local sector.
Section~\ref{sec:microscopic-origin} identifies the common one-particle character, explains the special role of the resolved conifold, and discusses
what is required to extend the construction to more general local
Calabi-Yau threefolds.  The appendices collect our conventions and the detailed transform and residue calculations.

\FloatBarrier
% ===========================================================================
\section{Non-critical M-theory: fillings and thermal spectral data}
\label{sec:ncm}

In this section, we collect the necessary ingredients of non-critical M-theory (NCM) for later discussion. For more details on NCM, we refer to\cite{Horava:2005tt,Horava:2005wm}. 
Subsection~\ref{sec:vacuum-map} distinguishes the relevant zero-temperature
fillings and derives their universal spectral kernels, while
subsection~\ref{sec:thermal} reviews the finite-temperature
Ho\v{r}ava-Keeler correspondence.

\subsection{Zero-temperature fillings and spectral characters}
\label{sec:vacuum-map}

Non-critial M-theory is the double-scaled theory of nonrelativistic fermions on a two-dimensional eigenvalue plane \cite{Horava:2005tt}. After the nonuniversal stabilizing terms are scaled away, its universal one-particle Hamiltonian is\begin{equation}
 h=\tfrac12\left(p_1^2+p_2^2\right)
   -\tfrac12\omega_0^2\left(\lambda_1^2+\lambda_2^2\right),
 \label{eq:h}
\end{equation}
where $\omega_0>0$ sets the oscillator scale and is kept explicit throughout.
With $\lambda_1=r\cos\theta$ and $\lambda_2=r\sin\theta$, the conserved angular
momentum is
\begin{equation}
 J=\lambda_1p_2-\lambda_2p_1=-\ii\partial_\theta,
 \qquad [h,J]=0.
 \label{eq:J}
\end{equation}
In a sector of fixed angular momentum $J=q$, the radial Hamiltonian is
\begin{equation}
 h_q=-\tfrac12\frac{\dd^2}{\dd r^2}
      -\tfrac12\omega_0^2r^2
      +\frac{q^2-\tfrac14}{2r^2},
 \qquad q\in\mathbb Z.
 \label{eq:0A-radial}
\end{equation}
The wavefunction $\psi_q(r,\theta)=e^{\ii q\theta}r^{-1/2}u_q(r)$ removes the
radial measure; its boundary condition is inherited from regular planar
wavefunctions.  Equation~\eqref{eq:0A-radial} is the type~0A matrix-model
Hamiltonian with $q$ units of signed RR flux
\cite{Douglas:2003up,Horava:2005tt}.

Place the Fermi edge at $-\mu$, with $\mu>0$.  Since the inverted oscillator is unbounded below, the filled Fermi sea extends to arbitrarily large radius and contains an infinite volume of deeply occupied states.  We therefore regulate the system by a rotationally invariant wall at large radius, with the corresponding cutoff denoted by $\Lambda$. The $\Lambda$-dependent contribution is nonuniversal and local. We use $E$ for an eigenvalue of $h$, and specify a zero-temperature vacuum by the occupation rule below the Fermi edge.

\begin{subequations}
\label{eq:zero-temperature-fillings}
\paragraph{Two-dimensional string vacua.}
Filling a single angular-momentum sector gives
\begin{equation}
 n_{q'}^{0A}(E)=\delta_{q'q}\,\theta(-\mu-E),
 \label{eq:occ-0A}
\end{equation}
which defines $|0A,q,\mu\rangle$, with all sectors $q'\neq q$ empty.  Its radial
dynamics is the type~0A theory at RR flux $q$.  Alternatively, the Cartesian
factorization $h=h_{\rm IHO}^{(1)}+h_{\rm IHO}^{(2)}$ permits one to fix the
energy and parity of one oscillator and fill the other.  A two-sided filling
of the remaining oscillator realizes type~0B; the conventional one-sided,
metastable filling gives the bosonic $c=1$ string \cite{Horava:2005tt,Douglas:2003up}.
These are different fillings of the same one-particle theory, not alternative
descriptions of the all-sector state used below.  Their topological B-model
relations are likewise distinct from the resolved-conifold A-model studied
here.\footnote{For example, compactifying the fixed-flux type~0A theory at
$R_{0A}=\sqrt{\alpha'/2}$ relates its perturbative sector to the B-model on
a deformed $\mathbb Z_2$ orbifold of the conifold
\cite{Hyun:2005aa}.  This special-radius compactification is additional
to selecting $J=q$ and is not the all-sector thermal correspondence.}

\paragraph{Natural M-theory vacuum.}
The rotationally invariant state $|M,\mu\rangle$ fills every angular-momentum
sector to the same energy,
\begin{equation}
 n_q^M(E)=\theta(-\mu-E),
 \qquad q\in\mathbb Z.
 \label{eq:occ-M}
\end{equation}
This planar Fermi sea, rather than a fixed-flux type~0A vacuum, enters the
unrefined resolved-conifold correspondence.
\end{subequations}

The corresponding one-particle Hilbert space is
$\mathcal H_1=\mathcal H_{\rm IHO}^{(1)}\otimes\mathcal H_{\rm IHO}^{(2)}$.
For $\tau>0$, its universal real-time character follows from the continued
Mehler trace:
\begin{equation}
 \begin{aligned}
 \chi_{\rm IHO}^{(i)}(\tau)
 &\equiv\operatorname{Tr}_{\mathcal H_{\rm IHO}^{(i)}}^{\rm reg}
 e^{-\ii\tau h_{\rm IHO}^{(i)}}
 =\sum_{n\geq0}e^{-\omega_0\tau(n+1/2)}
 =\frac{1}{2\sinh(\omega_0\tau/2)},\\
 \chi_0(\tau)
 &\equiv\operatorname{Tr}_{\mathcal H_1}^{\rm reg}e^{-\ii\tau h}
 =\prod_{i=1}^{2}\chi_{\rm IHO}^{(i)}(\tau)
 =\frac{1}{4\sinh^2(\omega_0\tau/2)}.
 \end{aligned}
 \label{eq:IHO-character}
\end{equation}
Here $\mathrm{reg}$ denotes the oscillatory Mehler trace in the double-scaled
resolvent prescription, not a trace-class thermal operator.  The sum is a
representation of the continued character, not a sum over discrete,
normalizable inverted-oscillator eigenstates.  In particular, $\tau$ is
real Schwinger proper time; the many-fermion thermal trace is introduced
only in subsection~\ref{sec:thermal}.  Appendix~\ref{app:HK-derivation}
evaluates the Mehler trace directly and records its short-time subtraction.

It is convenient to label the density by minus the physical energy,
$E=-\xi$, so that the Fermi edge occurs at $\xi=\mu$.  The resolvent identity
$(h+\xi-\ii0)^{-1}=\ii\int_0^\infty\dd\tau\,
 e^{-\ii\tau(h+\xi-\ii0)}$ gives
\begin{equation}
 \begin{aligned}
 \rho_0(\xi)
 &\equiv\frac{1}{\pi}\operatorname{Im}
 \operatorname{Tr}_{\mathcal H_1}^{\rm reg}\frac{1}{h+\xi-\ii0}
 =C_0(\Lambda)+\frac{1}{\pi}\operatorname{FP}\operatorname{Re}
 \int_0^\infty\dd\tau\,e^{-\ii\xi\tau}\chi_0(\tau)\\
 &=-\frac{\xi}{2\omega_0^2}\coth\!\left(\frac{\pi\xi}{\omega_0}\right)
   +\frac{\Lambda}{2\omega_0^2}.
 \end{aligned}
 \label{eq:rho0}
\end{equation}
The finite part removes the short-time divergence; the remaining local
constant is normalized as $C_0(\Lambda)=\Lambda/(2\omega_0^2)$ in the
Ho\v{r}ava--Keeler convention \cite{Horava:2005tt,Horava:2005wm}.
Thus \eqref{eq:rho0} is a double-scaled
spectral formula, with cutoff-suppressed terms omitted, rather than the
exact spectrum at a finite wall.  The subtracted universal part need not
be positive by itself; the regulator term is part of the physical density.

\paragraph{Rotating M-theory vacuum.}
Because $[h,J]=0$, the all-sector filling admits the stationary rotating
family studied by Ho\v{r}ava and Keeler \cite{Horava:2005tt}, with Fermi
surface
\begin{equation}
 K_\Omega\equiv h+\Omega J,
 \qquad K_\Omega=-\mu
 \quad\Longleftrightarrow\quad
 \mu_q=\mu+\Omega q,
 \qquad n_q^{\rm rot}(E)=\theta(-\mu_q-E).
 \label{eq:mu-shift}
\end{equation}
For real $\Omega$, this fills eigenstates of the Hermitian operator
$K_\Omega$; the sector-dependent $\mu_q$ need not be positive.  At fixed $q$,
the continued radial character is
$e^{-|q|\omega_0\tau}/[2\sinh(\omega_0\tau)]$.  Including the angular weight
and summing over sectors yields
\begin{equation}
 \begin{aligned}
 \chi_\Omega(\tau)
\equiv\operatorname{Tr}_{\mathcal H_1}^{\rm reg}e^{-\ii\tau K_\Omega}
  &=\frac{1}{2\sinh(\omega_0\tau)}
   \sum_{q\in\mathbb Z}e^{-|q|\omega_0\tau-\ii q\Omega\tau}\\
 &=\frac{1}{2[\cosh(\omega_0\tau)-\cos(\Omega\tau)]}\\
  &=\sum_{n_+,n_-\geq0}
    e^{-\tau[\alpha_+(n_++1/2)+\alpha_-(n_-+1/2)]}
   =\frac{e^{-\omega_0\tau}}
    {(1-e^{-\alpha_+\tau})(1-e^{-\alpha_-\tau})},
 \end{aligned}
 \label{eq:rotating-character}
\end{equation}
where $\alpha_\pm=\omega_0\pm\ii\Omega$.  The second line uses
$\sum_q e^{-|q|x-\ii qy}=\sinh x/(\cosh x-\cos y)$.
For the last line, the confining-oscillator labels obey
$q=n_+-n_-$ and $2n+|q|=n_++n_-$, with $n=\min(n_+,n_-)$.  They organize
the continued character, not a discrete spectrum of normalizable inverted-oscillator
states.  This circular form will isolate the orbital degrees of freedom in
section~\ref{sec:microscopic-character}.
Now $\xi$ labels minus an eigenvalue of $K_\Omega$, rather than of $h$.
The same resolvent transform gives
\begin{equation}
 \rho_\Omega(\xi)
 =C_\Omega(\Lambda)+\frac{1}{2\pi}\operatorname{FP}
 \int_0^\infty\dd\tau\,
 \frac{\cos(\xi\tau)}{\cosh(\omega_0\tau)-\cos(\Omega\tau)}.
 \label{eq:rhoOmega}
\end{equation}
This reproduces the rotating density of \cite{Horava:2005tt}, with the
short-time prescription made explicit.  Here $C_\Omega$ is independent of
$\xi$, and we write
$\rho_\Omega^{\rm univ}\equiv\rho_\Omega-C_\Omega$ for the finite part.
At $\Omega=0$, \eqref{eq:rhoOmega} reduces to \eqref{eq:rho0} with the same
regulator convention.

The factorization
\begin{equation}
 \cosh(\omega_0\tau)-\cos(\Omega\tau)
 =2\sinh\!\left(\frac{\alpha_+\tau}{2}\right)
   \sinh\!\left(\frac{\alpha_-\tau}{2}\right),
 \qquad \alpha_\pm\equiv\omega_0\pm\ii\Omega,
 \label{eq:factorisation}
\end{equation}
puts the density in the form
\begin{equation}
 \rho_\Omega(\xi)
 =C_\Omega(\Lambda)+\frac{1}{4\pi}\operatorname{FP}
 \int_0^\infty\dd\tau\,
 \frac{\cos(\xi\tau)}
 {\sinh(\alpha_+\tau/2)\sinh(\alpha_-\tau/2)}.
 \label{eq:rhoOmega-split}
\end{equation}
The two weights later identified with refinement are therefore already
present in the one-particle trace.  For $\tau>0$, the angular sum converges
in the strip
\begin{equation}
 |\operatorname{Im}\Omega|<\omega_0.
 \label{eq:strip}
\end{equation}
Within this strip the cosine integral also defines the analytic continuation
in $\Omega$.  At the purely imaginary boundary points
$\Omega=\pm\ii\omega_0$, one weight vanishes.  These are singular limits,
whose suitably rescaled free energies give the two Nekrasov--Shatashvili
limits, not additional finite-density vacua.

\subsection{The unrotated NCM ensemble and the perturbative A-model correspondence}
\label{sec:thermal}

Compactify Euclidean time on a circle of radius $R$ and set
$\beta_T=2\pi R$.  We reserve $\beta_T$ for inverse temperature, to distinguish
it from the curve class $\beta$ on the A-model side.  For the unrotated vacuum,
let $H$ be the second-quantized Hamiltonian and $\widehat N$ the fermion-number
operator.  Since the physical chemical potential is $-\mu$ and the
one-particle energies are $-\xi_a$, the grand-canonical trace is
\begin{equation}
 \begin{aligned}
 Z_{\NCM}(R,\mu)
 &=\operatorname{Tr}_{\mathcal F}e^{-\beta_T(H+\mu\widehat N)}
 =\prod_a\left(1+e^{\beta_T(\xi_a-\mu)}\right),\\
 \Gamma_{\NCM}(R,\mu)
 &\equiv\frac{1}{\beta_T}\log Z_{\NCM}(R,\mu)
 \longrightarrow\frac{1}{\beta_T}\int_{-\infty}^{\infty}\dd\xi\,
 \rho_0(\xi)\log\!\left(1+e^{\beta_T(\xi-\mu)}\right).
 \end{aligned}
 \label{eq:Gamma-thermal}
\end{equation}
Here $\mathcal F$ is fermionic Fock space and the arrow denotes the regulated
double-scaled limit.  The occupation tends to $\theta(\xi-\mu)$ as
$\beta_T\to\infty$, reproducing the filling at $E<-\mu$.
Our $\Gamma_{\NCM}=\beta_T^{-1}\log Z_{\NCM}$ is minus the conventional grand
potential.  With our energy convention, reversing the Fermi exponent, as in
the Ho\v{r}ava--Keeler orientation, gives the particle--hole conjugate trace.
With a common finite spectral regulator the two logarithms differ only by a
polynomial of degree one in $\mu$, as shown in
appendix~\ref{app:HK-derivation}.

The one-particle character and the many-body determinant thus play different
roles: $\chi_0$ determines $\rho_0$, while the Fermi factor supplies its thermal
occupation.  No high-temperature approximation has yet been made.  Taking
three $\mu$-derivatives eliminates the regulator constant and gives
\begin{equation}
 \partial_\mu^3\log Z_{\NCM}
 =-\frac14\int_0^\infty\dd\tau\;
 \frac{\tau^2\sin(\mu\tau)}
 {\sinh(\tau/2R)\sinh^2(\omega_0\tau/2)}.
 \label{eq:HK-third}
\end{equation}
The new factor $1/\sinh(\tau/2R)$ is the Fourier transform of the thermal
Fermi kernel; the other two factors come from \eqref{eq:IHO-character}.
Appendix~\ref{app:HK-derivation} gives the differentiation and transform,
including their signs and normalization.  Setting $\tau=R\sigma$ makes the
dimensionless combinations $R\mu$ and $R\omega_0$ explicit:
\begin{equation}
 \partial_\mu^{3}\log Z_{\NCM}
 =-\frac{R^3}{4}\int_0^\infty\dd\sigma\,
 \frac{\sigma^2\sin(R\mu\sigma)}
 {\sinh(\sigma/2)\sinh^2(R\omega_0\sigma/2)}.
 \label{eq:HK-third-scaled-main}
\end{equation}

The A-model variables and curve fugacity are
\begin{equation}
 \lambda=2\pi R\omega_0,
 \qquad t=2\pi R\mu+\ii\pi,
 \qquad Q\equiv e^{-t}=-e^{-2\pi R\mu}.
 \label{eq:HK-dictionary}
\end{equation}
The resolved conifold
$\mathcal O(-1)\oplus\mathcal O(-1)\to\mathbb P^1$ has one primitive
spin-zero curve multiplet.  Its perturbative nonconstant BPS free energy is
\cite{Gopakumar:1998ii,Gopakumar:1998jq}
\begin{equation}
 \cF^{\rm GV}_{\rm rc,BPS}(\lambda,t)
 =\sum_{k\ge1}\frac{e^{-kt}}{k\,[2\sin(k\lambda/2)]^2}.
 \label{eq:review-rc-GV}
\end{equation}
At fixed $\omega_0$, the weak-coupling expansion corresponds to
\begin{equation}
 R\to0,\qquad\mu\to\infty,\qquad R\mu\ \text{fixed}
 \quad\Longleftrightarrow\quad
 \lambda\to0,\qquad t\ \text{fixed}.
 \label{eq:HK-scaled-variables}
\end{equation}
Expanding the oscillator factor in \eqref{eq:HK-third-scaled-main}, evaluating
the thermal moments, and integrating three times gives
\begin{equation}
 [\log Z_{\NCM}]_{\rm nonlocal}
 \sim\cF^{\rm GV}_{\rm rc,BPS}(\lambda,t).
 \label{eq:HK-termwise}
\end{equation}
Here the subscript retains the nonpolynomial curve contribution after the
local polynomial and modulus-independent normalization have been removed.
The A-model pure-D0/constant-map sector defined in the introduction is
outside this comparison.  The polylogarithmic coefficients and their resummation into
\eqref{eq:review-rc-GV} are derived in appendix~\ref{app:HK-derivation}.
The symbol $\sim$ denotes equality of formal weak-coupling expansions;
it does not fix terms exponentially small in $1/\lambda$.  Recovering those
terms requires the unexpanded proper-time cycle, which is the subject of
section~\ref{sec:exact-completion}.

% ===========================================================================
\section{The rotating NCM ensemble and the refined A-model correspondence}
\label{sec:refinement}

The rotating ensemble is obtained by replacing $H$ in the Fock-space trace
by $H+\Omega\widehat J$.  Equivalently, one replaces $\rho_0$ by $\rho_\Omega$
in \eqref{eq:Gamma-thermal}.  Since $\xi$ now labels minus an eigenvalue of
$K_\Omega=h+\Omega J$, the angular shift is already included in the density
and the Fermi factor still contains the global $\mu$.  Expressing the same
ensemble sector by sector instead gives the depths $\mu_q=\mu+\Omega q$
of \eqref{eq:mu-shift}.  Applying the thermal transform to
\eqref{eq:rhoOmega-split} yields
\begin{equation}
 \partial_\mu^3\log Z_\NCM^{\rm rot}
 =-\frac14\int_0^\infty\dd\tau\;\operatorname{Im}\,
 \frac{\tau^2 e^{\ii\mu\tau}}
 {\sinh\!\left(\dfrac{\tau}{2R}\right)
  \sinh\!\left(\dfrac{\alpha_+\tau}{2}\right)
  \sinh\!\left(\dfrac{\alpha_-\tau}{2}\right)} .
 \label{eq:master-third}
\end{equation}
For real $\mu$ and physical real $\Omega$, the product of the oscillator
factors is real, so the imaginary part acts only on the exponential.
Rotation splits the two oscillator periods but leaves the thermal circle
unchanged.  The same real-projected expression holds on the positive-real
refinement slice specified below; continuation away from these slices is
performed on the cosine or sine integral, not on the nonholomorphic
operation $\operatorname{Im}$ itself.

To interpret the split, recall the protected M-theory index on
$X\times TN\times S^1$.  Transport around the circle rotates the two
complex planes of the transverse Taub--NUT space by equivariant parameters
$\eps_1$ and $\eps_2$.  A compensating $U(1)_R$ twist preserves the relevant
supercharge; on the self-dual locus $\eps_1=-\eps_2$ the index reduces to the
ordinary A-model \cite{Aganagic:2012hs}.  Its spin fugacities are
$\epsilon_{L,R}=(\eps_1\mp\eps_2)/2$, with conventions reviewed in
appendix~\ref{app:topstring}.

For the resolved conifold, the only primitive nonconstant BPS multiplet has
$(j_L,j_R)=(0,0)$.  Its spin characters are therefore trivial and the
perturbative refined free energy is \cite{Iqbal:2007ii,Chuang:2025aaa}
\begin{equation}
 \cF^{\rm ref}_{\rm rc,BPS,pert}(t;\eps_1,\eps_2)
 =\sum_{k\ge1}\frac{e^{-kt}}
 {4k\,\sin(k\eps_1/2)\sin(-k\eps_2/2)} .
 \label{eq:ref-GV}
\end{equation}
The denominator is the two-weight orbital contribution of the wrapped
particle.  It is the continuation of the circular-oscillator character
\eqref{eq:rotating-character}, as made explicit in
\eqref{eq:mic-character-factorization}; no intrinsic spin factor is
needed for this multiplet.  Matching these weights to the NCM periods gives
\begin{equation}
 \eps_1=2\pi R\alpha_+,\qquad
 -\eps_2=2\pi R\alpha_-,\qquad
 \lambda=2\pi R\sqrt{\alpha_+\alpha_-},\qquad
 \bpar^2=\frac{\alpha_+}{\alpha_-}.
 \label{eq:dictionary}
\end{equation}
We choose the square roots continuously from $\Omega=0$, so that
$\bpar=\alpha_+/\sqrt{\alpha_+\alpha_-}$,
$\eps_1=\lambda\bpar$ and $-\eps_2=\lambda/\bpar$.
This defines the common coupling through $\lambda^2=-\eps_1\eps_2$;
its unrefined value is $\lambda=2\pi R\omega_0$.  The K\"ahler parameter
remains $t=2\pi R\mu+\ii\pi$.

The map has two useful real slices:
\begin{equation*}
 \begin{aligned}
  \Omega\in\mathbb R:
  &\quad \alpha_-=\overline{\alpha_+},\qquad
  -\eps_2=\overline{\eps_1},\qquad |\bpar|=1,\\[1mm]
  \Omega=\ii\nu,\quad |\nu|<\omega_0:
  &\quad \alpha_+=\omega_0-\nu>0,\qquad
  \alpha_-=\omega_0+\nu>0,\qquad \bpar>0.
 \end{aligned}
\end{equation*}
Physical rotation therefore realizes the refined correspondence directly
on a complex-equivariant slice, with
$\lambda=2\pi R\sqrt{\omega_0^2+\Omega^2}$.
The continuation $\Omega=\ii\nu$ is useful because both oscillator periods
become positive and their proper-time poles lie on the same ray.  Note that it is an
analytic continuation of the spectral formula, not a second Hermitian rotating ensemble.  Section~\ref{sec:common-contour} describes the accompanying
contour deformation.  At $\nu=\pm\omega_0$, one of $\eps_1,-\eps_2$ vanishes;
the corresponding NS limit requires multiplication by that vanishing
parameter before taking the limit \cite{Nekrasov:2009rc}.

The perturbative matching can be established before deforming any contour.
Take $R\to0$ with $R\mu$, $\omega_0$ and $\Omega$ fixed, equivalently
$\lambda\to0$ at fixed $t$ and $\bpar$.  Expanding the two hyperbolic factors
in \eqref{eq:master-third} and evaluating the same thermal moments as in
section~\ref{sec:thermal} gives
\begin{equation}
 [\log Z_{\NCM}^{\rm rot}]_{\rm nonlocal}
 \sim\cF^{\rm ref}_{\rm rc,BPS,pert}(t;\eps_1,\eps_2).
 \label{eq:rot-refined-pert}
\end{equation}
For example, the first three orders are
\begin{equation*}
 \begin{aligned}
 \cF^{\rm ref}_{\rm rc,BPS,pert}
 \sim{}&\frac{\Li_3(Q)}{\lambda^2}
 +\frac{\bpar^2+\bpar^{-2}}{24}\Li_1(Q)\\
 &+\frac{7(\bpar^4+\bpar^{-4})+10}{5760}
   \lambda^2\Li_{-1}(Q)+O(\lambda^4),
 \qquad Q=e^{-t}.
 \end{aligned}
\end{equation*}
At $\bpar=1$ these reduce to the unrefined coefficients.
Appendix~\ref{app:HK-derivation} derives the general bivariate coefficient
and verifies that its formal resummation gives the two-sine denominator
in \eqref{eq:ref-GV}.  Rotation thus acts as refinement at the level of the
spectral character: the angular chemical potential resolves the two
oscillator weights in exactly the combination required by the protected
wrapped-M2 determinant. 
%===========================================================================
\section{Exact non-perturbative completion from non-critical M-theory}
\label{sec:exact-completion}

Sections~\ref{sec:thermal} and \ref{sec:refinement} matched the perturbative
weak-coupling expansions between NCM and unrefined and refined A-model on a resolved conifold, respectively.  We now keep the finite-$R$ NCM proper-time integral
unexpanded and show that its resolvent cycle yields the Hattab--Palti completion at
$\Omega=0$ and Chuang's refined completion for $\Omega\neq0$
\cite{Hattab:2024chf,Chuang:2025aaa}.

\subsection{The common Schwinger contour}
\label{sec:common-contour}

Equation~\eqref{eq:master-third} is the real proper-time representation from which
the complex contour will be derived.  Denote its integrand by
\begin{equation*}
 \mathcal K_\Omega(\tau)=
 \frac{\tau^2e^{\ii\mu\tau}}
 {\sinh(\tau/2R)\sinh(\alpha_+\tau/2)\sinh(\alpha_-\tau/2)} .
\end{equation*}
Its denominator is real on the real $\tau$ axis, both for physical rotation
$\Omega\in\mathbb R$ and on the positive-real slice $\Omega=\ii\nu$ of
section~\ref{sec:refinement}.  With the same symmetric prescription at $\tau=0$, this gives
\begin{equation}
 \mathcal K_\Omega(-\tau)=-\overline{\mathcal K_\Omega(\tau)},
 \qquad
 \int_0^\infty\!\operatorname{Im}\,\mathcal K_\Omega\,\dd\tau
 =\frac{1}{2\ii}\,\mathrm{PV}\!\int_{-\infty}^{\infty}
 \mathcal K_\Omega\,\dd\tau .
 \label{eq:half-to-full}
\end{equation}
Begin on the positive-real refinement slice.  For $\mu>0$, the factor
$e^{\ii\mu\tau}$ permits the two oriented real half-lines in
\eqref{eq:half-to-full} to be rotated towards the positive imaginary axis,
approaching its pole line from opposite sides.  Under $\tau=2\pi\ii Ru$,
the thermal dictionary gives
\[
 \frac{e^{-tu}}{1-e^{-2\pi\ii u}}
 =\frac{e^{-(t-\ii\pi)u}}{2\ii\sin(\pi u)},
 \qquad t-\ii\pi=2\pi R\mu.
\]
Together with \eqref{eq:dictionary}, this yields
\begin{equation}
 \begin{aligned}
 -\frac{1}{8\ii}\,\mathcal K_\Omega(\tau)\,\dd\tau
 &=(2\pi R)^3\partial_t^3
 \bigl[\cI(u;t,\eps_1,\eps_2)\,\dd u\bigr],\\
 \underbrace{\cI(u;t,\eps_1,\eps_2)\,\dd u}_{\displaystyle
 \omega_{\rm Sch}(u;t,\eps_1,\eps_2)}
 &\equiv\frac{\dd u}{u}\,\frac{e^{-tu}}{1-e^{-2\pi\ii u}}\,
 \frac{1}{4\sin(\eps_1u/2)\sin(-\eps_2u/2)} .
 \end{aligned}
 \label{eq:I-ref}
\end{equation}
We call $\omega_{\rm Sch}=\cI\,\dd u$ the BPS Schwinger one-form.  Its four
factors are the proper-time measure, the wrapped-M2 weight, the KK/D0 sum, and the
equivariant transverse determinant.  Here the KK/D0 sum is the tower attached to
the nonzero-curve-charge wrapped-M2 state; it is not the separate $\beta=0$
pure-D0 sector defined in the introduction.  The last factor is
$-\chi_\Omega(2\pi\ii Ru)$, so \eqref{eq:I-ref} already exhibits the
character factorization interpreted in
section~\ref{sec:microscopic-character}.  The contour derivation below
supplies the other ingredient needed for a microscopic spectral origin.

The inherited orientations of the two rotated half-lines give the nonzero-pole cycle
\begin{equation}
 \cC=-\int_{0^+}^{\infty e^{\ii0^+}}
     +\int_{0^+}^{\infty e^{\ii0^-}} ,
 \label{eq:contour}
\end{equation}
with the closing arc at infinity taken along a sequence avoiding the poles.
Its contribution vanishes on the thermal branch for $\mu>0$, and more generally
in the corresponding decay chamber.  The contour runs counterclockwise around
the nonzero poles between the two rays.  The notation $0^+$ excludes $u=0$:
the two lateral paths have a common positive starting point, and their
origin limit is taken only after they are combined.  The individual
undifferentiated ray integrals need not converge at the origin.  In the
unrefined limit this is the $0^+$ contour derived by Hattab and Palti from the
complexified Schwinger integral \cite{Hattab:2024chf}.

There is nevertheless an additional endpoint contribution on the NCM side.  During
the deformation from \eqref{eq:half-to-full}, the two real half-lines sweep two small
quarter-circles around $\tau=0$.  These arcs add rather than cancel and together form
half of a positively oriented loop around $u=0$.  This statement must be implemented
at the level of the three-derivative identity.  Indeed,
$\mathcal J(u)\equiv\partial_t^3\cI(u)=(-u)^3\cI(u)$ has only a simple pole at the
origin, even though $\cI$ itself has a fourth-order pole.  If
$\gamma_0(\rho)$ denotes the union of the two arcs at radius $\rho$, its total
angular opening is $\pi$, and hence
\begin{equation}
 \begin{aligned}
 \lim_{\rho\to0}\int_{\gamma_0(\rho)}\mathcal J(u)\,\dd u
 &=\ii\pi\,\Res_{u=0}\mathcal J(u)\\
 &=\partial_t^3P_{1/2}(t;\eps_1,\eps_2),
 \qquad
 P_{1/2}(t;\eps_1,\eps_2)
 \equiv\ii\pi\,\Res_{u=0}\cI(u;t,\eps_1,\eps_2).
 \end{aligned}
 \label{eq:P-half-def}
\end{equation}
The factor $1/2$ is therefore fixed by the angular opening of the endpoint contour,
not by any evenness property of $\cI$.  At the three-derivative level the complete
Wick-rotated NCM cycle is $\gamma_0+\cC$.  The endpoint identity fixes
$\partial_t^3P_{1/2}$; the residue in \eqref{eq:P-half-def} selects a canonical
antiderivative.  Any other choice differs by a polynomial of degree at most two in
$t$, precisely the integration polynomial introduced below.
Consequently, \eqref{eq:master-third} becomes the exact contour identity
\begin{equation}
 \partial_\mu^3\log Z_\NCM^{\rm rot}
 =(2\pi R)^3\partial_t^3
 \left[P_{1/2}(t;\eps_1,\eps_2)
 +\oint_{\cC}\cI(u;t,\eps_1,\eps_2)\,\dd u\right].
 \label{eq:main}
\end{equation}
Three integrations introduce a polynomial
$P_2(\mu;R,\Omega,\Lambda)$ of degree at most two in $\mu$, so that with
$P_{\rm loc}^{\NCM}\equiv P_{1/2}+P_2$ one has the finite-$R$ counterpart of the
perturbative decomposition of section~\ref{sec:thermal},
\begin{equation}
 \log Z_\NCM^{\rm rot}
 =P_{\rm loc}^{\NCM}
 +\cF_{\NCM}^{\rm BPS,np}(t;\eps_1,\eps_2),
 \qquad
 \cF_{\NCM}^{\rm BPS,np}
 \equiv\oint_{\cC}\cI(u;t,\eps_1,\eps_2)\,\dd u .
 \label{eq:integrated-main}
\end{equation}

Thus $\cF_{\NCM}^{\rm BPS,np}$ contains only the nonzero-pole residues;
$P_{1/2}$ is the canonical finite endpoint representative, while $P_2$ is the
independent integration polynomial.  This is also the separation needed for
comparison with Hattab--Palti and Chuang, whose $0^+$ BPS contours exclude the
origin.  We first establish the unrefined and refined nonzero-pole identities,
then evaluate the endpoint polynomial and compare normalizations in
section~\ref{sec:endpoint}.

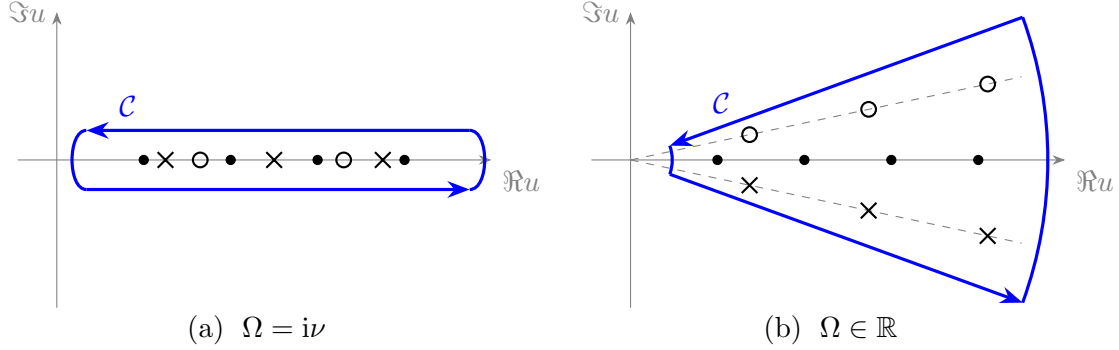
\begin{figure}[H]
\centering
\begin{tikzpicture}[scale=1.15,>=Stealth]

\begin{scope}
 \draw[gray,->] (-0.45,0) -- (5.0,0) node[below right]{\small $\Re u$};
 \draw[gray,->] (0,-1.7) -- (0,1.7) node[left]{\small $\Im u$};
 \foreach \x in {1,2,3,4}{\filldraw (\x,0) circle (1.5pt);}
 \foreach \x in {1.25,2.50,3.75}{%
   \draw[thick] (\x-0.09,-0.09)--(\x+0.09,0.09) (\x-0.09,0.09)--(\x+0.09,-0.09);}
 \foreach \x in {1.65,3.30}{\draw[thick] (\x,0) circle (2.4pt);}
 \draw[very thick,blue,->] (0.34,-0.34) -- (4.75,-0.34);
 \draw[very thick,blue,->] (4.75,0.34) -- (0.34,0.34);
 \draw[very thick,blue] (0.34,0.34) .. controls (0.12,0.34) and (0.12,-0.34) .. (0.34,-0.34);
 \draw[very thick,blue] (4.75,-0.34) .. controls (4.98,-0.34) and (4.98,0.34) .. (4.75,0.34);
 \node[blue] at (0.82,0.62) {\small $\cC$};
 \node at (2.3,-1.95) {\small (a)\; $\Omega=\ii\nu$};
\end{scope}

\begin{scope}[xshift=6.6cm]
 \draw[gray,->] (-0.45,0) -- (5.0,0) node[below right]{\small $\Re u$};
 \draw[gray,->] (0,-1.7) -- (0,1.7) node[left]{\small $\Im u$};
 \foreach \x in {1,2,3,4}{\filldraw (\x,0) circle (1.5pt);}
 \draw[dashed,gray] (0,0) -- (12:4.6);
 \draw[dashed,gray] (0,0) -- (-12:4.6);
 \foreach \r in {1.4,2.8,4.2}{%
   \draw[thick] ({\r*cos(-12)-0.09},{\r*sin(-12)-0.09})--({\r*cos(-12)+0.09},{\r*sin(-12)+0.09});
   \draw[thick] ({\r*cos(-12)-0.09},{\r*sin(-12)+0.09})--({\r*cos(-12)+0.09},{\r*sin(-12)-0.09});}
 \foreach \r in {1.4,2.8,4.2}{\draw[thick] ({\r*cos(12)},{\r*sin(12)}) circle (2.4pt);}
 \draw[very thick,blue,->] (-20:0.48) -- (-20:4.8);
 \draw[very thick,blue,->] (20:4.8) -- (20:0.48);
 \draw[very thick,blue] (20:0.48) arc (20:-20:0.48);
 \draw[very thick,blue] (-20:4.8) arc (-20:20:4.8);
 \node[blue] at (1.05,0.66) {\small $\cC$};
 \node at (2.3,-1.95) {\small (b)\; $\Omega\in\mathbb R$};
\end{scope}

\end{tikzpicture}
\caption{Pole lattices of \eqref{eq:I-ref}.  Filled dots denote the integer KK/D0
poles, crosses the $\eps_1$ family, and open circles the $-\eps_2$ family.  The two
oscillator lattices are real for $\Omega=\ii\nu$ and form conjugate rays for physical
real rotation.  Blue contours schematically enclose the nonzero poles; their
inner connectors exclude $u=0$.  The separate half-origin contribution
$P_{1/2}$ is not drawn.}
\label{fig:poles}
\end{figure}

For $\Omega=\ii\nu$ the three pole lattices lie on the positive real $u$ axis.  For
real $\Omega$ the integer lattice stays there while the two oscillator lattices move
onto conjugate rays, and the contour deforms with them, as in
figure~\ref{fig:poles}; at commensurate periods coincident poles are evaluated as a
single higher-order pole.

\FloatBarrier

\subsection{The unrotated NCM ensemble and the unrefined completion}
\label{sec:contour-unref}

For the non-rotating vacuum, $\Omega=0$ and $\alpha_+=\alpha_-=\omega_0$, so that
$\eps_1=-\eps_2=\lambda=2\pi R\omega_0$.  The Schwinger form \eqref{eq:I-ref} becomes
\begin{equation}
 \cI_{\rm unref}(u)
 =\frac{1}{u}\,
 \frac{e^{-tu}}{1-e^{-2\pi\ii u}}\,
 \frac{1}{[2\sin(\lambda u/2)]^2}.
 \label{eq:I-unref}
\end{equation}
Because $u=0$ has already been separated, $\cC$ receives contributions only from two
nonzero pole families.  Assume first that $\lambda/\pi$ is irrational, so that they
do not collide.  Separate residue sums are understood first in a chamber where they
converge and elsewhere by analytic continuation.  The KK/D0 factor has simple poles
at $u=k\in\mathbb Z_{>0}$, and
\begin{equation}
 2\pi\ii\,\Res_{u=k}\cI_{\rm unref}(u)
 =\frac{e^{-kt}}{k[2\sin(k\lambda/2)]^2}.
 \label{eq:unref-integer-residue}
\end{equation}
Their sum is exactly the perturbative resolved-conifold multicover series
\eqref{eq:review-rc-GV}.

The second family lies at
$u=u_\ell\equiv2\pi\ell/\lambda$, $\ell\in\mathbb Z_{>0}$.  These are double
poles because the two oscillator sine factors coincide.  To evaluate their contribution, define
\begin{equation*}
 g(u)\equiv\frac{e^{-tu}}{u(1-e^{-2\pi\ii u})}.
\end{equation*}
Since
\begin{equation*}
 \cI_{\rm unref}(u)
 =\frac{1}{\lambda^2}
 \left[\frac{g(u_\ell)}{(u-u_\ell)^2}
       +\frac{g'(u_\ell)}{u-u_\ell}+O(1)\right],
\end{equation*}
the residue is $g'(u_\ell)/\lambda^2$.  The useful observation is
\[
 \frac{\ii}{\ell}g(u_\ell)
 =\frac{\lambda}{4\pi\ell^2}
   \frac{e^{-2\pi\ell(t-\ii\pi)/\lambda}}
        {\sin(2\pi^2\ell/\lambda)},
 \qquad
 \left.\frac{\partial u_\ell}{\partial\lambda}\right|_t
 =-\frac{2\pi\ell}{\lambda^2}.
\]
The chain rule therefore gives the termwise identity
\begin{equation}
 2\pi\ii\,\Res_{u=u_\ell}\cI_{\rm unref}(u)
 =-\left.\frac{\partial}{\partial\lambda}\right|_{t}
 \left[
  \frac{\lambda}{4\pi\ell^2}\,
  \frac{\exp\!\left[-\dfrac{2\pi\ell}{\lambda}
       (t-\ii\pi)\right]}
       {\sin(2\pi^2\ell/\lambda)}
 \right].
 \label{eq:unref-double-residue}
\end{equation}
The derivative is taken at fixed $t$; the NCM relation
$t=2\pi R\mu+\ii\pi$ is imposed only after the residue has been evaluated.
Thus both the derivative and the factor $\ell^{-2}$ follow from the double
pole.  Its explicit Laurent coefficient is recorded in
appendix~\ref{app:unrefined-residues}.  To sum these terms, introduce the
resolved-conifold Nekrasov--Shatashvili free energy in the Hattab--Palti normalization,
\begin{equation*}
 F_{\rm NS}^{\rm rc}(\hbar,T)
 \equiv\frac{1}{4\pi}\sum_{\ell\geq1}
 \frac{e^{-\ell T}}{\ell^2\sin(\pi\ell\hbar)}.
\end{equation*}
Summing \eqref{eq:unref-integer-residue} and
\eqref{eq:unref-double-residue} therefore yields
\begin{equation}
 \begin{aligned}
 \oint_{\cC}\cI_{\rm unref}(u)\,\dd u
 &=\cF_{\rm rc,BPS}^{\rm GV}(\lambda,t)
 -\left.\frac{\partial}{\partial\lambda}\right|_{t}
 \left\{\lambda F_{\rm NS}^{\rm rc}\!\left(
 \frac{2\pi}{\lambda},
 \frac{2\pi}{\lambda}(t-\ii\pi)\right)\right\}\\
 &\equiv\cF_{\rm HP}^{\rm rc,BPS,np}(t,\lambda).
 \end{aligned}
 \label{eq:HP-NS}
\end{equation}
The second line is precisely the Hattab--Palti nonzero-pole completion of the resolved
conifold \cite[eqs.~(2.14)--(2.16)]{Hattab:2024chf}.  In particular, $F_{\rm NS}^{\rm rc}$ has not been
postulated as an independent NCM quantity: equation~\eqref{eq:unref-double-residue}
shows directly that it is the compact notation for the oscillator-pole residue sum.
The same GV-plus-NS structure also appears in Alim's earlier
nonperturbative conifold construction \cite{Alim:2021intrinsic}.

The distinction between the two pole families is also the perturbative versus
nonperturbative distinction.  The oscillator residues contain
\begin{equation}
 \left|\exp\!\left[-\frac{2\pi\ell}{\lambda}
 (t-\ii\pi)\right]\right|
 =\exp\!\left[-\frac{2\pi\ell\,\operatorname{Re}t}{\lambda}\right]
 =e^{-2\pi\ell\mu/\omega_0}
 \quad\text{on }\quad
 t=2\pi R\mu+\ii\pi,\ \lambda=2\pi R\omega_0 .
 \label{eq:np-scale}
\end{equation}
At fixed $t$ with $\operatorname{Re}t>0$ this is
$e^{-\mathrm{const}/\lambda}$, beyond all orders of the weak-coupling expansion
as $\lambda\to0^+$.  The genus expansion generated by the integer poles does
not fix the coefficient with which this sector enters the completion.  It is included because the NCM-derived
cycle encloses the oscillator poles.  Combining \eqref{eq:integrated-main} and
\eqref{eq:HP-NS} gives the main unrefined statement in its simplest form,
\begin{equation}
 \log Z_\NCM^{\Omega=0}
 =P_{1/2}+P_2+\cF_{\rm HP}^{\rm rc,BPS,np}.
 \label{eq:NCM-HP-completion}
\end{equation}
Thus NCM reproduces the Hattab--Palti completion through the same Schwinger one-form
and the same $0^+$ nonzero-pole contour, while independently selecting the half-origin
term discussed in section~\ref{sec:endpoint}.  If the two nonzero lattices collide at
rational $\lambda/\pi$, the separately written sums are singular and must be
replaced by the combined higher-order residues of the original contour; the contour
integral itself remains the primary definition.

\subsection{The rotating NCM ensemble and the refined completion}
\label{sec:contour-refined}

For $\Omega\neq0$ the two oscillator periods split.  On the positive-real refinement
slice, $\eps_1>0$ and $-\eps_2>0$, the nonzero poles enclosed by $\cC$ form the three
families
\begin{equation}
 u=k,\qquad
 u=u_\ell^{(1)}\equiv\frac{2\pi\ell}{\eps_1},\qquad
 u=u_\ell^{(2)}\equiv-\frac{2\pi\ell}{\eps_2},
 \qquad k,\ell\in\mathbb Z_{>0}.
 \label{eq:refined-pole-families}
\end{equation}
We first take the three periods to be noncommensurate and work in a chamber where the
separate residue sums converge; the resulting expressions are continued analytically
elsewhere.  At an integer pole only the KK/D0 denominator vanishes, and hence
\begin{equation}
 2\pi\ii\,\Res_{u=k}\cI(u;t,\eps_1,\eps_2)
 =\frac{e^{-kt}}
 {4k\sin(k\eps_1/2)\sin(-k\eps_2/2)}.
 \label{eq:refined-integer-residue}
\end{equation}
Summing \eqref{eq:refined-integer-residue} gives exactly the perturbative refined GV
series \eqref{eq:ref-GV}.

At either oscillator pole one of the equivariant sine factors vanishes instead.  For
generic noncommensurate periods the poles are simple, and direct evaluation gives
\begin{equation}
 \begin{aligned}
 2\pi\ii\,\Res_{u=u_\ell^{(1)}}\cI
 &=\frac{\ii(-1)^\ell}{2\ell}\,
 \frac{e^{-2\pi\ell t/\eps_1}}
 {\sin(-\pi\ell\eps_2/\eps_1)
  \left(1-e^{-4\pi^2\ii\ell/\eps_1}\right)},\\[1mm]
 2\pi\ii\,\Res_{u=u_\ell^{(2)}}\cI
 &=\frac{\ii(-1)^\ell}{2\ell}\,
 \frac{e^{2\pi\ell t/\eps_2}}
 {\sin(-\pi\ell\eps_1/\eps_2)
  \left(1-e^{4\pi^2\ii\ell/\eps_2}\right)}.
 \end{aligned}
 \label{eq:res-np}
\end{equation}
At fixed $t$, these are beyond-all-orders sectors in the corresponding
weak-coupling decay chamber.  Their NCM scale is clearest after rewriting the
KK/D0 denominator as a sine, which shifts the exponential weight from $t$ to
$t-\ii\pi$; the algebra is given in appendix~\ref{app:np-res}.
On the positive-real slice the two shifted weights have magnitudes
$e^{-2\pi\ell\mu/\alpha_+}$ and $e^{-2\pi\ell\mu/\alpha_-}$.
For physical real $\Omega$, instead, they obey
\begin{equation*}
 \left|e^{-2\pi\ell(t-\ii\pi)/\eps_1}\right|
 =\left|e^{2\pi\ell(t-\ii\pi)/\eps_2}\right|
 =\exp\!\left[-\frac{2\pi\ell\mu\omega_0}
                    {\omega_0^2+\Omega^2}\right].
\end{equation*}
The complete oscillator residues then occur in conjugate pairs, as required
for a real grand potential.  This also makes explicit why the real-period
suppression formula cannot be read literally with complex $\alpha_\pm$.

Appendix~\ref{app:chuang} specializes Chuang's refined wrapped-M2 contour
\cite[eq.~(60)]{Chuang:2025aaa} to the unique spin-zero curve multiplet of
the resolved conifold.  The result is
precisely the one-form \eqref{eq:I-ref} integrated over \eqref{eq:contour}.
Combining that specialization with the three residue calculations above gives
\begin{equation}
 \begin{aligned}
 \log Z_\NCM^{\rm rot}-P_{1/2}-P_2
 &=\oint_{\cC}\cI(u;t,\eps_1,\eps_2)\,\dd u\\
 &=\cF_{\rm rc,BPS,pert}^{\rm ref}
 +\sum_{\ell\geq1}2\pi\ii
 \left(\Res_{u=u_\ell^{(1)}}\cI
      +\Res_{u=u_\ell^{(2)}}\cI\right)\\
 &\equiv\cF_{\rm Chuang}^{\rm rc,BPS,np}(t;\eps_1,\eps_2).
 \end{aligned}
 \label{eq:refined-BPS-identity}
\end{equation}
This is the refined analogue of \eqref{eq:NCM-HP-completion}: the integer residues
give the perturbative GV block, while the two oscillator families give the two
nonperturbative sectors, with their lateral prescription inherited from NCM.  For
physical real $\Omega$ the oscillator lattices and the contour continue to conjugate
rays.  When periods become commensurate, coincident poles are evaluated together.  In
particular, as $\Omega\to0$ the two simple oscillator families merge into the double
poles whose residues give \eqref{eq:unref-double-residue}.  Appendix~\ref{app:np-res}
contains the residue algebra and an equivalent three-frame resummation; the next
subsection extracts the associated Stokes jump.
\subsection{Stokes discontinuities and the quantum dilogarithm}
\label{sec:stokes}

The oscillator poles have two related but distinct roles.  In
\eqref{eq:refined-BPS-identity} they contribute to the fixed NCM completion;
a Stokes discontinuity instead compares two lateral prescriptions for the
same KK/D0 image.  To make this distinction explicit, expand the KK/D0 factor
on the two sides of the real axis, as in
\eqref{eq:KK-lateral-expansions}.  With
$\eps_1=\lambda\bpar$ and $-\eps_2=\lambda/\bpar$, the resulting image
integrals are
\begin{equation*}
 \cF_n^\pm=\int_{\gamma_\pm}\frac{\dd u}{u}
 \frac{e^{-A_nu/(2\pi)}}
 {4\sin(\lambda\bpar u/2)\sin(\lambda u/(2\bpar))},
 \qquad A_n=2\pi(t+2\pi\ii n),\quad n\in\mathbb Z,
\end{equation*}
where $\gamma_\pm$ pass above or below the positive oscillator poles and use
the same subtraction at the origin.  In the Borel variable
$s=\lambda(t+2\pi\ii n)u$, these poles lie at
$s=\ell A_n/\bpar$ and $s=\ell\bpar A_n$.  For $\bpar>0$ the two
families therefore belong to the same Stokes ray.  The change of variable
and the local subtractions are detailed in
appendix~\ref{app:stokes-derivation}.

Define the discontinuity by the lower-minus-upper convention,
$\operatorname{Disc}_{A_n}\cF\equiv\cF_n^--\cF_n^+$.
The difference contour is counterclockwise, so it equals $2\pi\ii$ times
the sum of the oscillator residues of this image, without the KK/D0 factor.
For noncoincident poles this gives
\begin{equation}
 \begin{aligned}
 \operatorname{Disc}_{A_n}\cF
 &=\ii\sum_{\ell\geq1}\frac{(-1)^\ell}{\ell}
 \left[
  \frac{e^{-\ell A_n/(\lambda\bpar)}}
       {2\sin(\pi\ell/\bpar^2)}
  +
  \frac{e^{-\ell\bpar A_n/\lambda}}
       {2\sin(\pi\ell\bpar^2)}
 \right]\\
 &=-\log\Phi_{\bpar}\!\left(-\frac{A_n}{2\pi\lambda}\right),
 \end{aligned}
 \label{eq:stokes}
\end{equation}
where the convention for the Faddeev quantum dilogarithm is fixed by
\eqref{eq:Faddeev-definition}.  Its integral has precisely these two pole
families, and its residue expansion fixes the sign in \eqref{eq:stokes}.
Since $N_{0,0}^{\beta_0}=1$, the result is the resolved-conifold specialization
of Chuang's Stokes automorphism \cite[eq.~(35)]{Chuang:2025aaa}:
\begin{equation}
 \frac{Z_n^-}{Z_n^+}
 =\Phi_{\bpar}\!\left(-\frac{A_n}{2\pi\lambda}\right)^{-1},
 \qquad Z_n^\pm\equiv e^{\cF_n^\pm}.
 \label{eq:stokes-multiplicative}
\end{equation}

The collinear derivation uses the positive-real refinement slice
$\Omega=\ii\nu$, $|\nu|<\omega_0$.  For physical real angular velocity,
the NCM dictionary instead gives
\begin{equation}
 \begin{aligned}
 \eps_1&=2\pi R(\omega_0+\ii\Omega),
 &-\eps_2&=2\pi R(\omega_0-\ii\Omega)=\overline{\eps_1},\\
 \lambda&=2\pi R\sqrt{\omega_0^2+\Omega^2},
 &\bpar^2&=\frac{\omega_0+\ii\Omega}{\omega_0-\ii\Omega},
 \qquad |\bpar|=1 .
 \end{aligned}
 \label{eq:physical-b-domain}
\end{equation}
The branch of $\bpar$ is chosen continuously from $\bpar=1$ at $\Omega=0$.
The oscillator lattices in the $u$ plane now lie on conjugate rays; in the
Borel plane the corresponding rays are directed by $A_n/\bpar$ and
$\bpar A_n$.  Continuing the contour with the poles gives the combined
change from crossing both rays with the above orientation,
\begin{equation}
 \boxed{
 \operatorname{Disc}^{(1+2)}_{A_n}\cF
 \equiv
 \operatorname{Disc}^{(1)}_{A_n/\bpar}\cF
 +\operatorname{Disc}^{(2)}_{\bpar A_n}\cF
 =-\log\Phi_{\bpar}\!\left(-\frac{A_n}{2\pi\lambda}\right) .
 }
 \label{eq:combined-two-ray-disc}
\end{equation}
Thus the quantum-dilogarithm factor continues to the physical $|\bpar|=1$
slice as a \emph{two-ray} Stokes transformation.  Crossing only one ray
retains only its corresponding residue family; it does not give the full
factor in \eqref{eq:combined-two-ray-disc}.  Nor does this transformation
assert a discontinuity of the equilibrium grand potential as real $\Omega$
is varied.  As $\Omega\to0$, the two families must be combined before taking
the limit.  Appendix~\ref{app:stokes-derivation} evaluates the resulting
double poles and recovers the unrefined dilogarithmic jump.

\subsection{The origin contribution and the remaining normalization}
\label{sec:endpoint}

The identities above concern only nonzero poles.  To recover the NCM grand
potential, one must also restore the endpoint contribution that is absent
from the $0^+$ BPS contours of Hattab--Palti and Chuang.  The symmetric NCM
Wick rotation gives the half-origin cycle of
\eqref{eq:P-half-def}, whose canonical finite part is
\begin{equation}
 P_{1/2}(t;\eps_1,\eps_2)
 \equiv
 \underset{\rho\to0}{\operatorname{FP}}
 \int_{\gamma_0(\rho)}\cI(u;t,\eps_1,\eps_2)\,\dd u
 =\ii\pi\operatorname{Res}_{u=0}\cI(u;t,\eps_1,\eps_2).
 \label{eq:P-half-finite-part}
\end{equation}
Here the finite part removes the negative powers of $\rho$ before the
origin limit is taken.  At the three-derivative level no such subtraction
is needed: $\partial_t^3\cI$ has only a simple pole, and the total angular
opening $\pi$ fixes its contribution directly.

Expanding the three factors of \eqref{eq:I-ref} near $u=0$ gives the compact
representative
\begin{equation*}
 P_{1/2}
 =\frac{(t-\ii\pi)^3+\pi^2(t-\ii\pi)}{12\eps_1\eps_2}
 +\frac{t-\ii\pi}{48}
  \left(\frac{\eps_1}{\eps_2}+\frac{\eps_2}{\eps_1}\right).
\end{equation*}
The expansion and its Barnes representation are derived in
appendix~\ref{app:transform}.  Its cubic coefficient passes a direct
spectral check:
\[
 \partial_\mu^3 P_{1/2}
 =-\frac{\pi R}{\alpha_+\alpha_-},
\]
which is the large-$\mu$ limit of the original proper-time integral.
This cubic term is fixed by the resolvent cycle.  The lower-degree terms
specify the displayed finite-part representative, not additional data
fixed by the third derivative; changing them can be compensated by the
integration polynomial $P_2$.

At $\eps_1=-\eps_2=\lambda$, this representative is exactly one half of
the Hattab--Palti zero-pole polynomial, as shown in
\eqref{eq:HP-zero-pole}.  The half-loop selected by NCM therefore agrees
with the single-chamber normalization discussed by Hattab and Palti, where
the full zero-pole term is shared by the two flop chambers
\cite{Hattab:2024chf}.
Combining the endpoint and nonzero-pole pieces yields
\begin{equation}
 \boxed{
 \log Z_{\NCM}^{\rm rot}
 =\cF_{\rm Chuang}^{\rm rc,BPS,np}(t;\eps_1,\eps_2)
 +P_{1/2}(t;\eps_1,\eps_2)+P_2,
 \qquad \deg_\mu P_2\leq2 .
 }
 \label{eq:refined-final-decomposition}
\end{equation}
At $\Omega=0$ the first term becomes the Hattab--Palti completion.
The residual $P_2$ records the local thermodynamic normalization lost upon
three differentiations.  A fully normalized topological-string free energy
also requires the $t$-independent pure-D0/constant-map sector, which the
nonconstant NCM block does not determine.  Thus the exact identification
fixes the nonconstant BPS completion and the cubic local term, while an
absolute partition-function equality requires these remaining normalization
data to be matched separately.

\section{Microscopic spectral origin and extensions beyond the conifold}
\label{sec:microscopic-origin}

Having identified the exact Hattab--Palti and Chuang completions in
section~\ref{sec:exact-completion}, we now turn to the microscopic reason why the
same Schwinger data emerge from non-critical M-theory.  The essential point is already
visible at the one-particle level.  The rotating all-$q$ vacuum resolves the planar
oscillator into two circular modes with frequencies
$\alpha_\pm=\omega_0\pm\ii\Omega$, while the resolved conifold has an exceptionally
simple nonconstant BPS spectrum: a single primitive spin-zero wrapped-M2 multiplet,
$N_{0,0}^{\beta_0}=1$ \cite{Gopakumar:1998jq,Iqbal:2007ii}.  Consequently no
additional intrinsic-spin factor is required, and the NCM orbital spectrum is enough
to reproduce the complete one-particle contribution of this state.  This observation
also makes clear which part of the construction is special to the conifold and which
part could plausibly survive for a more general Calabi--Yau.

\subsection{The conifold particle from the NCM spectrum}
\label{sec:microscopic-character}

Recall that the rotating one-particle trace \eqref{eq:rotating-character} factorizes
into the two circular-oscillator weights $\alpha_\pm$.  At finite temperature the same
trace enters the grand potential through \eqref{eq:master-third}; no new spectral
degrees of freedom are introduced in passing from the zero-temperature density to the
thermal ensemble.  After the symmetric contour rotation of
section~\ref{sec:common-contour}, equation~\eqref{eq:I-ref} may be rewritten directly
in terms of this one-particle trace as
\begin{equation}
\omega_{\rm Sch}(u)
=\frac{\dd u}{u},
\frac{e^{-tu}}{1-e^{-2\pi\ii u}},
\big[-\chi_\Omega(2\pi\ii Ru)\big].
\label{eq:mic-character-factorization}
\end{equation}
Using \eqref{eq:dictionary}, the factor in brackets is precisely
$[4\sin(\eps_1u/2)\sin(-\eps_2u/2)]^{-1}$, the equivariant orbital determinant
appearing in the refined Gopakumar--Vafa formula \eqref{eq:ref-GV}.  

Each factor in \eqref{eq:mic-character-factorization} has a direct interpretation.
The continued NCM trace supplies the two equivariant orbital modes; $e^{-tu}$ is the
weight of the primitive curve-charged state; and
$(1-e^{-2\pi\ii u})^{-1}$ generates its Kaluza--Klein, or D0, tower around the
M-theory circle.  Because the conifold multiplet has $(j_L,j_R)=(0,0)$, there is no
additional protected spin character multiplying these factors.  Thus the same
one-particle data that follow from the rotating NCM Hamiltonian are precisely those
entering the Schwinger determinant of the conifold wrapped-M2 state.  The
zero-curve-charge pure-D0 sector remains separate, as discussed in
sec.~\ref{sec:endpoint}.

This one-particle matching also clarifies where the nonperturbative information enters.  The determinant determines the proper-time integrand, while the NCM resolvent supplies the real integration cycle that accompanies it.  Under the symmetric contour deformation of sec.~\ref{sec:common-contour}, this cycle becomes the nonzero-pole contour \eqref{eq:contour}, together with the half-origin contribution \eqref{eq:P-half-def}; as shown in sec.~\ref{sec:exact-completion}, the resulting residues reproduce the Hattab--Palti and Chuang completions.  In this sense, the microscopic spectral interpretation involves not only the one-particle determinant but also the contour prescription inherited from the same NCM spectral problem.

%There is also a useful, though secondary, consistency check.  On the positive-real refinement slice, rescaling proper time by $s=\alpha_+\tau$ puts the two-period orbital factor into the finite-radius $c=1$ form, with $\mathcal R=\alpha_+/\alpha_-=\bpar^2$; exchanging the two circular modes sends $\mathcal R\to\mathcal R^{-1}$.  This is the familiar radius-deformed $c=1$ universality of the refined conifold \cite{Klebanov:1991string,Nakayama:2010aaa}. It concerns the perturbative conifold singularity and a common orbital kernel, not an equivalence between the rotating all-$q$ vacuum and the separate $c=1$ filling of NCM, and it is not used in the nonperturbative derivation above.

\subsection{Beyond the resolved conifold}
\label{sec:generic-CY}

The same reasoning also isolates what is special about the conifold.  The two-weight
orbital determinant is universal for a five-dimensional BPS particle in the
$\Omega$-background, whereas the spectrum of primitive curve charges and protected
spins depends on the Calabi--Yau geometry.  Chuang's general contour formula
\cite[eq.~(60)]{Chuang:2025aaa}, reviewed in appendix~\ref{app:chuang}, makes this
separation explicit.  Using the chamber representative $T_\beta$ defined there, the
geometry-dependent protected one-particle data can be collected into
\begin{equation}
\mathcal B_X(u)
=\sum_{\beta>0,j_L,j_R}N_{j_L,j_R}^{\beta},
e^{-uT_\beta},
\chi_{j_L}(e^{\ii u\epsilon_L})
\chi_{j_R}(e^{\ii u\epsilon_R}),
\qquad
\epsilon_{L,R}=\frac{\eps_1\mp\eps_2}{2}.
\label{eq:generic-BPS-character}
\end{equation}
The full Schwinger one-form is obtained from
\eqref{eq:mic-character-factorization} by replacing the conifold factor $e^{-tu}$
with $\mathcal B_X(u)$.  For the resolved conifold,
$\mathcal B_{\rm rc}(u)=e^{-tu}$, so the NCM orbital sector already exhausts the
nonconstant primitive one-particle data.  For a general geometry,
$\mathcal B_X$ contains precisely the additional information that is not present in
the planar oscillator.

This suggests, but does not determine, how a microscopic extension might look.  One
possibility would be to retain the NCM orbital sector and supplement it by an internal
supersymmetric quantum mechanics whose protected states carry the curve charges and
spins appearing in \eqref{eq:generic-BPS-character}.  Local $\mathbb P^2$ gives a
natural first example, because it has only one K\"ahler modulus but already possesses
nontrivial protected spin.  Its degree-one curves are lines, and the corresponding
refined BPS multiplet has $(j_L,j_R)=(0,1)$ with multiplicity one
\cite{Choi:2012jz}.  In the conventions above its contribution
would be
\begin{equation}
\mathcal B_{\mathbb P^2}^{(d=1)}(u)
=e^{-uT_H}\chi_1(e^{\ii u\epsilon_R})
=e^{-uT_H}\bigl[1+2\cos(2u\epsilon_R)\bigr],
\label{eq:localP2-degree-one-character}
\end{equation}
where $H$ is the line class.  On the unrefined locus this becomes
$3e^{-uT_H}$, but at generic refinement it is not equivalent to three spinless
conifold states; the nontrivial spin dependence has to come from additional protected
degrees of freedom.

Equation~\eqref{eq:localP2-degree-one-character} should therefore be regarded as a concrete first example, rather than evidence for a particular enlarged NCM model.  It would be
interesting to understand whether an internal quantum mechanics can be coupled to the
orbital sector in such a way that its primitive protected spectrum emerges
microscopically and its regulated trace admits a natural continuation of the NCM
proper-time cycle.  Geometries with several independent K\"ahler parameters would
raise the further question of how the corresponding charge chemical potentials are
realized.  Whether such an extension exists remains open. We hope we can come back to this problem in the future.

%===========================================================================

\section{Conclusions} 
\label{sec:conclusion} 
We have promoted the Ho\v{r}ava--Keeler finite-temperature correspondence from a comparison of weak-coupling expansions \cite{Horava:2005wm} to an equality of the nonperturbatively completed nonconstant sector of the resolved-conifold free energy. For the unrotated all-$q$ vacuum, the NCM proper-time integral reproduces the Hattab--Palti contour formula for the resolved conifold \cite{Hattab:2024chf, Hattab:2024ewk}. Turning on the angular chemical potential splits the oscillator periods according to \begin{equation*} \eps_1=2\pi R(\omega_0+\ii\Omega),\qquad -\eps_2=2\pi R(\omega_0-\ii\Omega), \end{equation*} and yields Chuang's refined completion \cite{Chuang:2025aaa}. In both cases the integer poles give the Gopakumar--Vafa multicovers, while the oscillator poles give the exponentially small sectors absent from the genus expansion. The microscopic content of this result lies in the NCM spectral problem. Its one-particle trace produces the complete two-weight orbital character of the unique spinless conifold curve multiplet, and its resolvent selects the real proper-time cycle. After contour rotation that cycle fixes the lateral prescription, including at commensurate periods where individual residue sums cease to be meaningful. This establishes a microscopic spectral origin for the Hattab--Palti and Chuang BPS completions, not an identification of NCM fermions with wrapped M2-branes. The symmetric rotation also fixes the half-origin representative discussed in section~\ref{sec:endpoint}; the residual degree-two thermodynamic polynomial and the pure-D0/constant-map normalization remain independent data.

The completed conifold free energy obtained here is consistent with earlier nonperturbative descriptions based on resurgence, exact Chern--Simons theory, and spectral-zeta methods \cite{Pasquetti:2009jg,Krefl:2015vna,Hatsuda:2015oaa}. Riemann--Hilbert and difference-equation approaches provide further exact
descriptions of the resolved conifold
\cite{Bridgeland:2017vbr,Alim:2021intrinsic}, while related resurgence,
exact WKB and quantum-curve analyses have been developed for the resolved
conifold and its mirror \cite{Alim:2022wkb}. In the refined theory, the two oscillator-pole families found here coincide with the $\bpar$- and $\bpar^{-1}$-rescaled Borel singularities, and the associated Stokes jump has the quantum-dilogarithm structure identified by Alexandrov, Marino, and Pioline \cite{Alexandrov:2023refined}.  Our result therefore does not introduce a different conifold transseries; rather, it explains how the same nonperturbative structure emerges from the independently defined NCM Hamiltonian and resolvent.  This perspective is reminiscent of the ABJM Fermi-gas description and its relation to refined topological strings \cite{Marino:2011eh,Hatsuda:2013oxa}, as well as the
later spectral formulation of topological strings in terms of quantized mirror curves \cite{Grassi:2014zfa}. The mechanisms are nevertheless different: in mirror-curve constructions the Calabi--Yau geometry is built into the quantum operator, whereas in NCM the resolved-conifold Schwinger structure emerges from the continuum inverted oscillator after thermal
compactification and contour rotation.

The conifold example also suggests where the present construction may cease to be sufficient.  Its unusually simple BPS spectrum allows the NCM orbital factor to account for the entire nonconstant one-particle contribution.  For a more general local Calabi--Yau, one expects additional information associated with primitive curve classes, protected multiplicities, and nontrivial spin representations.  It is therefore natural to ask whether the NCM spectral problem can be enlarged so that such data arise from additional microscopic degrees of freedom rather than being supplied externally.  Local $\mathbb P^2$ provides a particularly simple testing ground: it has a single K\"ahler modulus but already carries nontrivial degree and spin information, with \eqref{eq:localP2-degree-one-character} giving a first hint.  One possible direction would be to couple the NCM orbital sector to an appropriate BPS or Donaldson--Thomas quantum mechanics.  Whether such an extension can simultaneously reproduce the primitive protected spectrum and furnish a natural continuation of the NCM spectral cycle remains an open question.

\subsection*{Acknowledgements}
F.X. is supported by Beijing Natural
Science Foundation No.~JR26001. Portions of the literature cross-checking, algebraic verification, exposition, and importantly \LaTeX{} editing were largely assisted by OpenAI ChatGPT 5.6 and Anthropic Claude Opus 5. %The author independently reviewed the resulting arguments and calculations and takes full responsibility for the content of the paper.

\appendix

% ===========================================================================
\section{Topological-string and \texorpdfstring{$\Omega$}{Omega}-background conventions}
\label{app:topstring}

This appendix collects the topological-string background used in the body.  Local
polynomials and constant-map terms of the noncompact conifold are kept separate from
the nonconstant BPS sector throughout.

\subsection{The five-dimensional index, the \texorpdfstring{$U(1)_R$}{U(1)R} twist, and
\texorpdfstring{$\eps_{1,2}$}{epsilon1,2}}

M-theory on a Calabi--Yau threefold $X$ produces a five-dimensional theory whose
massive BPS particles include M2-branes wrapped on holomorphic curves.  Place this
theory on $TN\times S^1$, with complex coordinates $(z_1,z_2)$ on the two asymptotic
planes of the Taub--NUT space.  The refined A-model is the protected index in the
twisted identification
\begin{equation}
 (z_1,z_2,y)\sim
 (e^{\ii\eps_1}z_1,e^{\ii\eps_2}z_2,y+2\pi R_M).
 \label{eq:app-Omega-monodromy}
\end{equation}
Equivalently, with
$\mathfrak q=e^{\ii\eps_1}$ and
$\mathfrak t=e^{-\ii\eps_2}$, the index may be written
\cite{Aganagic:2012hs}
\begin{equation}
 Z_X^{\rm ref}
 =\mathrm{Tr}_{\mathcal H_{\rm BPS}}
 (-1)^F\,
 \mathfrak q^{\,S_1-S_R}\,
 \mathfrak t^{\,S_R-S_2}\,
 e^{-\boldsymbol t\cdot\widehat{\boldsymbol d}}.
 \label{eq:app-refined-index}
\end{equation}
Here $\widehat{\boldsymbol d}$ is the curve-charge operator; it is distinct
from the scalar fugacity $Q=e^{-t}$ used for the conifold.
The generators $S_1,S_2$ rotate the two complex planes and $S_R$ is the
$U(1)_R$ generator.  A generic two-plane rotation does not by itself preserve the
same supercharge; it must be accompanied by this $U(1)_R$ action on the
Calabi--Yau.  Thus the $R$-symmetry twist is not a third independent refinement
parameter: supersymmetry fixes it once $\eps_1$ and $\eps_2$ are chosen.  For the
noncompact toric geometries relevant here the required $U(1)$ action is available.

When $\mathfrak q=\mathfrak t$, equivalently $\eps_1=-\eps_2=\lambda$, the $R$-charge
drops out and the index reduces to the ordinary topological A-model.  Away from this
self-dual locus, the combinations
\begin{equation}
 \epsilon_L=\frac{\eps_1-\eps_2}{2},
 \qquad
 \epsilon_R=\frac{\eps_1+\eps_2}{2}
 \label{eq:app-epsilon-LR}
\end{equation}
couple to the two Cartans of the five-dimensional massive little group
$Spin(4)=SU(2)_L\times SU(2)_R$.  This is why refined BPS invariants retain both
spins $(j_L,j_R)$.

\subsection{Ordinary and refined GV expansions}

The closed topological-string free energy has the formal genus expansion
\begin{equation}
 \cF_X(\lambda,t)=\sum_{g\ge0}\lambda^{2g-2}\cF_{g,X}(t).
 \label{eq:app-genus-expansion}
\end{equation}
Its nonconstant part can be reorganized in terms of Gopakumar--Vafa
invariants \cite{Gopakumar:1998ii,Gopakumar:1998jq}:
\begin{equation}
 \cF_{X,\mathrm{BPS}}
 =\sum_{\beta>0}\sum_{g\ge0}n_\beta^g
 \sum_{k\ge1}\frac{1}{k}
 \left[2\sin\left(\frac{k\lambda}{2}\right)\right]^{2g-2}
 e^{-k\beta\cdot\boldsymbol t}.
 \label{eq:app-GV-general}
\end{equation}
For
$X_{\rm rc}=\mathcal O(-1)\oplus\mathcal O(-1)\to\mathbb P^1$ the only primitive
invariant is
\begin{equation}
 n_{\beta_0}^{0}=1,
 \qquad n_\beta^g=0\quad\text{otherwise in the nonconstant sector},
 \label{eq:app-GV-conifold}
\end{equation}
and hence
\begin{equation}
 \cF_{\rm rc,BPS}^{\rm unref}(t,\lambda)
 =\sum_{k\ge1}\frac{e^{-kt}}
 {k[2\sin(k\lambda/2)]^2}.
 \label{eq:app-conifold-unref}
\end{equation}

In the refined theory we use $N_{j_L,j_R}^{\beta}$ for the signed
coefficients of the protected index, not necessarily nonnegative spin
multiplicities.  Thus any conventional spin-statistics sign
$(-1)^{2j_L+2j_R}$ is included in $N_{j_L,j_R}^{\beta}$.
With $\chi_j(e^{\ii x})=\sum_{m=-j}^{j}e^{2\ii mx}$, the nonconstant
perturbative BPS free energy is \cite{Iqbal:2007ii,Chuang:2025aaa}
\begin{equation}
 \cF_{X,\mathrm{BPS}}^{\rm ref,pert}
 =\sum_{\beta>0,j_L,j_R}N_{j_L,j_R}^{\beta}
 \sum_{k\ge1}\frac{e^{-k\beta\cdot\boldsymbol t}}{k}
 \frac{\chi_{j_L}(e^{\ii k\epsilon_L})
       \chi_{j_R}(e^{\ii k\epsilon_R})}
 {4\sin(k\eps_1/2)\sin(-k\eps_2/2)} .
 \label{eq:app-refined-GV}
\end{equation}
For the resolved conifold
\begin{equation}
 N_{0,0}^{\beta_0}=1,
 \qquad N_{j_L,j_R}^{\beta}=0\quad\text{otherwise in the nonconstant sector},
 \label{eq:app-refined-conifold-invariant}
\end{equation}
so the characters are trivial and
\begin{equation}
 \cF_{\rm rc,BPS}^{\rm ref,pert}
 =\sum_{k\ge1}\frac{e^{-kt}}
 {4k\sin(k\eps_1/2)\sin(-k\eps_2/2)} .
 \label{eq:app-refined-conifold}
\end{equation}
At $\eps_1=-\eps_2$ this reduces to
\eqref{eq:app-conifold-unref}.

\subsection{Complex Schwinger contours}

The GV series fixes only the perturbative expansion.  Hattab and Palti retain the
complexified proper-time contour of the wrapped-M2 determinant
\cite{Hattab:2024chf,Hattab:2024ewk}.  In the fundamental B-field chamber their
resolved-conifold BPS block is
\begin{equation}
 \cF_{\rm HP}^{\rm rc}(t,\lambda)
 =\oint_{\cC}\frac{\dd u}{u}
 \frac{e^{-tu}}{1-e^{-2\pi\ii u}}
 \frac{1}{[2\sin(\lambda u/2)]^2},
 \qquad
 \cC=-\int_{0^+}^{\infty e^{\ii0^+}}
      +\int_{0^+}^{\infty e^{\ii0^-}} .
 \label{eq:app-HP-contour}
\end{equation}
The poles at $u=k$ reproduce the perturbative GV series; the poles
$u=2\pi\ell/\lambda$ produce its non-perturbative completion.  Their contour starts
at $0^+$ and excludes the origin.  Hattab and Palti therefore supplement the contour
by polynomial terms, and separately by the pure-D0 contribution, when specifying an
absolute free energy.  For the resolved conifold, restoring the full zero-pole residue
gives
\begin{equation}
 F_0^{\rm Poly}=-\frac{t^3}{6}+\frac{\ii\pi}{2}t^2+\frac{\pi^2}{3}t,
 \qquad
 F_1^{\rm Poly}=-\frac{t}{12}+\frac{\ii\pi}{12}.
 \label{eq:app-zero-polynomials}
\end{equation}
In the chamber normalization discussed by Hattab and Palti, this zero pole
is shared by the two flop chambers, so one chamber carries one half of
\eqref{eq:app-zero-polynomials}.

Chuang's refinement inserts the two equivariant determinants and the BPS characters
\cite{Chuang:2025aaa}.  Its proper-time rays also begin at $0^+$, so the refined
formula likewise fixes the nonzero-pole BPS block rather than the local origin
polynomial.  For the spin-zero conifold block it reduces to
\begin{equation}
 \cF_{\rm Chuang}^{\rm rc,BPS,np}(t;\eps_1,\eps_2)
 =\oint_{\cC}\frac{\dd u}{u}
 \frac{e^{-tu}}{1-e^{-2\pi\ii u}}
 \frac{1}{4\sin(\eps_1u/2)\sin(-\eps_2u/2)} .
 \label{eq:app-Chuang-conifold}
\end{equation}
Its three positive pole families are
\begin{equation}
 u=k,\qquad
 u=\frac{2\pi\ell}{\eps_1},\qquad
 u=\frac{2\pi\ell}{-\eps_2},
 \qquad k,\ell\ge1.
 \label{eq:app-three-poles}
\end{equation}
The first family is \eqref{eq:app-refined-conifold}; the other two are the refined
non-perturbative sectors.

% ===========================================================================
\section{Spectral transforms and the Ho\v{r}ava--Keeler expansion}
\label{app:HK-derivation}

We derive the continued character, its density transform, and the thermal
kernel before expanding in the A-model coupling.  This keeps the spectral
regularization separate from the asymptotic expansion.

\subsection{The Mehler trace and the density subtraction}

For real $\tau>0$, the diagonal one-dimensional inverted-oscillator
propagator is \cite[eq.~(3.25)]{Klebanov:1991string}
\begin{equation*}
 \langle x|e^{-\ii\tau h_{\rm IHO}}|x\rangle
 =\left(\frac{\omega_0}{2\pi\ii\sinh(\omega_0\tau)}\right)^{1/2}
 \exp\!\left[\ii\omega_0x^2\tanh(\omega_0\tau/2)\right].
\end{equation*}
Taking its Fresnel integral, with a Gaussian regulator removed from the
convergent side, gives $1/[2\sinh(\omega_0\tau/2)]$.
The planar trace is its square.  More generally, inserting the rotation
$e^{-\ii\Omega\tau J}$ in the planar Mehler kernel gives
\begin{equation*}
 \begin{aligned}
 \chi_\Omega(\tau)
 &=\lim_{\eta\to0^+}
 \frac{\omega_0}{2\pi\ii\sinh(\omega_0\tau)}
 \int_{\mathbb R^2}\dd^2x\,
 \exp\!\left[
 -\eta|x|^2+\frac{\ii\omega_0[\cosh(\omega_0\tau)-\cos(\Omega\tau)]}
 {\sinh(\omega_0\tau)}|x|^2\right]\\
 &=\frac{1}{2[\cosh(\omega_0\tau)-\cos(\Omega\tau)]}.
 \end{aligned}
\end{equation*}
This independently reproduces the angular-sector sum
\eqref{eq:rotating-character}.  Both calculations initially apply for real
$\Omega$; the resulting kernel is analytic in the strip \eqref{eq:strip}.

At small $\tau$, $\chi_\Omega(\tau)=
1/(\alpha_+\alpha_-\tau^2)+O(1)$.  For real $\xi$, the finite part used in
the density is therefore
\begin{equation*}
 \rho_\Omega^{\rm univ}(\xi)
 =\frac{1}{\pi}\lim_{\delta\to0^+}
 \left[\int_\delta^\infty\dd\tau\,
 \cos(\xi\tau)\chi_\Omega(\tau)
 -\frac{1}{\alpha_+\alpha_-\delta}\right].
\end{equation*}
The local regulator term $C_\Omega(\Lambda)$ is added separately.
At zero rotation, direct integration gives
$\rho_0^{\rm univ}(0)=-1/(2\pi\omega_0)$, while the convergent difference is
\begin{equation*}
 \begin{aligned}
 \rho_0^{\rm univ}(\xi)-\rho_0^{\rm univ}(0)
 &=\frac{1}{4\pi}\int_0^\infty\dd\tau\,
 \frac{\cos(\xi\tau)-1}{\sinh^2(\omega_0\tau/2)}\\
 &=-\frac{\xi^2}{\pi\omega_0}
 \sum_{n\geq1}\frac{1}{n^2\omega_0^2+\xi^2}
 =-\frac{\xi}{2\omega_0^2}\coth(\pi\xi/\omega_0)
  +\frac{1}{2\pi\omega_0}.
 \end{aligned}
\end{equation*}
Adding $C_0(\Lambda)$ reproduces \eqref{eq:rho0}, including its finite
value at $\xi=0$.

\subsection{The Fermi factor and the exact thermal transform}

The physical filling at energy below $-\mu$ gives the exponent
$\beta_T(\xi-\mu)$ in \eqref{eq:Gamma-thermal}.  The opposite, hole-oriented
exponent differs by the identity
\begin{equation*}
 \log(1+e^{\beta_T(\xi-\mu)})
 -\log(1+e^{\beta_T(\mu-\xi)})=\beta_T(\xi-\mu).
\end{equation*}
For a common finite spectral regulator, the sum of the right-hand side is
affine in $\mu$.  Its coefficients can diverge as the regulator is removed,
but it has zero third derivative and belongs to the local integration
polynomial.  The two orientations therefore give the same spectral
identity used in the main text, although their first derivatives count
particles and holes differently.

With $\beta_T=2\pi R$, define
$f_{\beta_T}(x)=(1+e^{\beta_T x})^{-1}$ and
$w_{\beta_T}(x)=f_{\beta_T}(x)[1-f_{\beta_T}(x)]
=1/[4\cosh^2(\beta_T x/2)]$.  Differentiating the physical trace gives
\begin{equation}
 \partial_\mu\Gamma_{\NCM}
 =-\int\dd\xi\,\rho_0(\xi)f_{\beta_T}(\mu-\xi),
 \qquad
 \partial_\mu^2\Gamma_{\NCM}
 =\beta_T\int\dd\xi\,\rho_0(\xi)w_{\beta_T}(\xi-\mu).
 \label{eq:B-derivatives}
\end{equation}
Since $w_{\beta_T}$ is even, one more derivative and an integration by
parts give
\begin{equation}
 \partial_\mu^3\log Z_{\NCM}
 =\beta_T^2\int_{-\infty}^{\infty}\dd\xi\,
 \rho_0'(\xi)w_{\beta_T}(\xi-\mu).
 \label{eq:B-third}
\end{equation}
The boundary term vanishes when the spectral cutoff is removed at fixed
$R,\mu$, because the Fermi kernel decays exponentially.
The constant $C_0(\Lambda)$ drops out of $\rho_0'$.  The two required
transforms are
\begin{equation}
 \begin{aligned}
 \rho_0'(\xi)
 &=-\frac{1}{4\pi}\int_0^\infty\dd\tau\,
 \frac{\tau\sin(\xi\tau)}{\sinh^2(\omega_0\tau/2)},\\
 \int_{-\infty}^{\infty}\dd\xi\,
 e^{-\ii\xi\tau}w_{\beta_T}(\xi-\mu)
 &=\frac{\tau e^{-\ii\mu\tau}}
 {4\pi R^2\sinh(\tau/2R)}.
 \end{aligned}
 \label{eq:B-transforms}
\end{equation}
The second follows from
$\int\dd y\,e^{-\ii ky}\operatorname{sech}^2y=\pi k/\sinh(\pi k/2)$
with $y=\beta_T(\xi-\mu)/2$.  The first is already convergent at the
origin: its numerator is $O(\tau^2)$.  Moreover, the bound
$|\sin(\xi\tau)|\leq|\xi\tau|$ and the finite first moment of
$w_{\beta_T}(\xi-\mu)$ justify exchanging the two integrals.
Inserting \eqref{eq:B-transforms} into \eqref{eq:B-third} gives
\eqref{eq:HK-third}, with prefactor $-1/4$.
If written instead as the imaginary part of a complex exponential,
that projection must be taken before the origin limit: the unprojected
complex kernel has a simple pole there.

\subsection{The unrefined weak-coupling expansion}

Write $\partial_{R\mu}$ for differentiation with respect to the dimensionless
combination $R\mu$.  After $\tau=R\sigma$,
\begin{equation}
 \partial_{R\mu}^{3}\log Z_{\NCM}
 =-\frac14\int_0^\infty\dd\sigma\,
 \frac{\sigma^2\sin(R\mu\sigma)}
 {\sinh(\sigma/2)\sinh^2(R\omega_0\sigma/2)} .
 \label{eq:HK-third-scaled}
\end{equation}
Expanding the oscillator factor with
\begin{equation}
 \frac{1}{\sinh^2(x/2)}
 =-4\sum_{g\ge0}\frac{(2g-1)B_{2g}}{(2g)!}\,x^{2g-2},
 \qquad 0<|x|<2\pi,
 \label{eq:csch-Bernoulli}
\end{equation}
the required moments can be evaluated without a contour prescription.  For real
$R\mu$, the absolutely convergent geometric expansion gives
\begin{align*}
 \frac{1}{\sinh(\sigma/2)}
 &=2\sum_{n=0}^{\infty}e^{-(n+1/2)\sigma},\\
 I_0(R\mu)
 &\equiv\int_0^\infty\dd\sigma\,
 \frac{\sin(R\mu\sigma)}{\sinh(\sigma/2)}\\
 &=2R\mu\sum_{n=0}^{\infty}
 \frac{1}{(n+1/2)^2+(R\mu)^2}
 =\pi\tanh(\pi R\mu),
\end{align*}
where the last equality is the Mittag--Leffler expansion of $\tanh$.  Differentiation
under the integral sign is then justified for every $g\in\mathbb Z_{\ge0}$ and yields
\begin{equation*}
 \int_0^\infty\dd\sigma\,
 \frac{\sigma^{2g}\sin(R\mu\sigma)}{\sinh(\sigma/2)}
 =(-1)^g\partial_{R\mu}^{2g}I_0(R\mu)
 =(-1)^g\pi\partial_{R\mu}^{2g}
 \tanh(\pi R\mu).
\end{equation*}
The same identity extends analytically to
$|\operatorname{Im}(R\mu)|<\tfrac12$.  Using these moments, one obtains
\begin{equation}
 \partial_{R\mu}^{3}\log Z_{\NCM}
 \sim-\pi\sum_{g\ge0}
 \frac{(2g-1)B_{2g}}{(2g)!}(\ii R\omega_0)^{2g-2}
 \partial_{R\mu}^{2g}\tanh(\pi R\mu),
 \label{eq:HK-derivative-series}
\end{equation}
in agreement with the Ho\v{r}ava--Keeler expansion \cite{Horava:2005wm}.
The Laurent expansion is not uniform over the full $\sigma$ integration
range.  It gives an asymptotic expansion after integration, not a convergent
exchange of infinite sums; the exponentially small sectors require the
unexpanded contour of section~\ref{sec:exact-completion}.

With $Q=-e^{-2\pi R\mu}=e^{-t}$ one has
$\tanh(\pi R\mu)=1+2\Li_0(Q)$ and
$\partial_{R\mu}\Li_s(Q)=-2\pi\Li_{s-1}(Q)$.  Hence every term of
\eqref{eq:HK-derivative-series} has an explicit antiderivative,
\begin{equation}
 \begin{aligned}
 -\frac{\pi}{R^2\omega_0^2}\tanh(\pi R\mu)
 &=\partial_{R\mu}^{3}
 \left[-\frac{\pi(R\mu)^3}{6R^2\omega_0^2}
       +\frac{\Li_3(Q)}{\lambda^2}\right],\\[1mm]
 -\pi\frac{(2g-1)B_{2g}}{(2g)!}(\ii R\omega_0)^{2g-2}
 \partial_{R\mu}^{2g}\tanh(\pi R\mu)
 &=\partial_{R\mu}^{3}\left[
 \frac{(-1)^{g-1}B_{2g}}{2g(2g-2)!}
 \lambda^{2g-2}\Li_{3-2g}(Q)\right],
 \qquad g\ge1.
 \end{aligned}
 \label{eq:HK-antiderivatives}
\end{equation}
Here $(2g-1)/(2g)!=1/[2g(2g-2)!]$ and
$(\ii R\omega_0)^{2g-2}=(-1)^{g-1}\lambda^{2g-2}/(2\pi)^{2g-2}$.
Consequently,
\begin{equation}
 \log Z_{\NCM}\sim P_{\rm loc}
 +\frac{\Li_3(Q)}{\lambda^2}+\frac{\Li_1(Q)}{12}
 +\sum_{g\ge2}\frac{(-1)^{g-1}B_{2g}}{2g(2g-2)!}
 \lambda^{2g-2}\Li_{3-2g}(Q),
 \label{eq:HK-full-series}
\end{equation}
where the local part may be written without introducing an additional scaled
chemical potential as
\begin{equation}
 P_{\rm loc}=P_2^{\rm HK}(\mu;R,\omega_0,\Lambda)
 -\frac{(t-\ii\pi)^3}{12\lambda^2}
 -\frac{t-\ii\pi}{24},
 \qquad \deg_\mu P_2^{\rm HK}\le2 .
 \label{eq:HK-local-sector}
\end{equation}
Equation \eqref{eq:HK-full-series} is the Ho\v{r}ava--Keeler result written in
the positive Schwinger variable $\lambda$ introduced in \eqref{eq:intro-HK-map};
in particular, $\Li_1(Q)=-\log(1+e^{-2\pi R\mu})$.  The cubic in
\eqref{eq:HK-local-sector} comes from the $g=0$ antiderivative, while $P_2^{\rm HK}$
is the degree-$\leq2$ integration ambiguity.  The independent contour
calculation supplies a compatible representative: specializing \eqref{eq:P-half} to
$\eps_1=-\eps_2=\lambda$ yields
\begin{equation}
 P_{1/2}(t;\lambda,-\lambda)
 =-\frac{(t-\ii\pi)^3}{12\lambda^2}
 -\frac{\pi^2(t-\ii\pi)}{12\lambda^2}
 -\frac{t-\ii\pi}{24},
 \label{eq:HK-endpoint-crosscheck}
\end{equation}
whose cubic and displayed genus-one term agree with
\eqref{eq:HK-local-sector}.  The remaining linear term relates the two
integration-polynomial conventions by
\begin{equation*}
 P_2=P_2^{\rm HK}+\frac{\pi^2(t-\ii\pi)}{12\lambda^2},
\end{equation*}
where $P_2$ is defined relative to $P_{1/2}$ in
section~\ref{sec:exact-completion}.  Only the cubic is fixed by the
three-derivative identity; the genus-one linear term is part of the chosen
local representative.

Removing $P_{\rm loc}$ and inserting
$\Li_{3-2g}(Q)=\sum_{k\ge1}k^{2g-3}Q^k$, the coefficient of $Q^k$ in
\eqref{eq:HK-full-series} is $1/\{k[2\sin(k\lambda/2)]^2\}$ by
\begin{equation}
 \frac{1}{[2\sin(x/2)]^2}=\frac{1}{x^2}+\frac{1}{12}
 +\sum_{g\ge2}\frac{(-1)^{g-1}B_{2g}}{2g(2g-2)!}x^{2g-2}
 \label{eq:sine-Bernoulli}
\end{equation}
at $x=k\lambda$, which establishes \eqref{eq:HK-termwise}; the sense in which
the interchange of the two sums is formal was discussed below that equation.
Ho\v{r}ava and Keeler state the comparison after removing the normalization
ambiguity at a reference modulus,
\begin{equation}
 \frac{Z_A(\lambda,t)}{Z_A(\lambda,t_*)}
 \sim\frac{Z_{\NCM}(R,\mu)}{Z_{\NCM}(R,\mu_*)},
 \qquad
 t_*=2\pi R\mu_*+\ii\pi .
 \label{eq:HK-normalized}
\end{equation}
The ratio removes a modulus-independent constant at each order in $\lambda$, but
it does not remove a degree-two polynomial; \eqref{eq:HK-normalized} therefore
presupposes that the local polynomials on the two sides have first been fixed to
a common representative.

\subsection{The refined expansion and its NS boundary}

Applying the same thermal transform to the split density
\eqref{eq:rhoOmega-split}, then setting $\tau=R\sigma$, gives
\begin{equation}
 \partial_{R\mu}^3\log Z_\NCM^{\rm rot}
 =-\frac14\int_0^\infty\dd\sigma\,
 \frac{\sigma^2\sin(R\mu\sigma)}
 {\sinh(\sigma/2)\sinh(R\alpha_+\sigma/2)\sinh(R\alpha_-\sigma/2)},
 \label{eq:rot-scaled}
\end{equation}
valid on both the physical real-$\Omega$ slice and the positive-real
refinement slice.  Its continuation uses the sine representation as
explained in section~\ref{sec:refinement}.  The two-frequency Laurent
coefficients factorize,
\begin{equation}
 \begin{aligned}
 \frac{1}{4\sinh(x/2)\sinh(y/2)}
 &=\sum_{m,n\ge0}a_ma_n\,x^{2m-1}y^{2n-1},\\
 a_m&=\frac{B_{2m}(1/2)}{(2m)!}
 =\frac{(2^{1-2m}-1)B_{2m}}{(2m)!}.
 \end{aligned}
 \label{eq:bivariate-csch}
\end{equation}
This Laurent series is valid for $0<|x|<2\pi$ and $0<|y|<2\pi$.  Since
$x/[2\sinh(x/2)]$ generates the Bernoulli polynomials at $1/2$, one has $a_0=1$,
$a_1=-1/24$ and $a_2=7/5760$.  The thermal factor $1/\sinh(\sigma/2)$ is untouched, so
the moment identity applies term by term and, with $\eps_1=2\pi R\alpha_+$ and
$-\eps_2=2\pi R\alpha_-$,
\begin{equation}
 [\log Z_\NCM^{\rm rot}]_{\rm nonlocal}
 \sim\sum_{m,n\ge0}a_ma_n(-1)^{m+n}
 \eps_1^{2m-1}(-\eps_2)^{2n-1}\Li_{3-2m-2n}(Q).
 \label{eq:rot-bivariate-polylog}
\end{equation}
Before the local terms are removed, the $(m,n)=(0,0)$ contribution to
\eqref{eq:rot-scaled} is
\begin{equation*}
 -\frac{4\pi^3\tanh(\pi R\mu)}{\eps_1(-\eps_2)},
\end{equation*}
whose constant part integrates to
\begin{equation*}
 -\frac{2\pi^3(R\mu)^3}{3\,\eps_1(-\eps_2)}
 =-\frac{2\pi R\mu^3}{12\,\alpha_+\alpha_-},
\end{equation*}
the refined classical term obtained independently in \eqref{eq:large-mu} from the
small-$\tau$ region of the exact kernel.

The formal resummation follows from analytic continuation of the Laurent kernel.  Since $\ii^{2m-1}\ii^{2n-1}=(-1)^{m+n-1}$, the
alternating sign in \eqref{eq:rot-bivariate-polylog} is precisely what evaluating
\eqref{eq:bivariate-csch} at imaginary argument produces, and
\begin{equation}
 \begin{aligned}
 \sum_{m,n\ge0}a_ma_n(-1)^{m+n}(k\eps_1)^{2m-1}(-k\eps_2)^{2n-1}
 &=-\frac{1}{4\sinh(\ii k\eps_1/2)\sinh(-\ii k\eps_2/2)}\\[1mm]
 &=\frac{1}{4\sin(k\eps_1/2)\sin(-k\eps_2/2)} .
 \end{aligned}
 \label{eq:bivariate-continuation}
\end{equation}
Inserting $\Li_{3-2m-2n}(Q)=\sum_{k\ge1}k^{2m+2n-3}Q^k$ into
\eqref{eq:rot-bivariate-polylog} therefore gives \eqref{eq:rot-refined-pert}.  As in
the unrefined case the resummation is formal, the Laurent expansion at fixed $k$ being
valid only for $|k\eps_1|<2\pi$ and $|k\eps_2|<2\pi$.  At commensurate
periods the completed contour, rather than separately singular residue
families, supplies the definition.

For completeness, the NS boundary is a scaled limit.  Term by term in
$Q$, or in a convergence chamber before analytic continuation,
\begin{equation*}
 \lim_{-\eps_2\to0^+}(-\eps_2)
 \cF^{\rm ref}_{\rm rc,BPS,pert}(t;\eps_1,\eps_2)
 =\sum_{k\geq1}\frac{Q^k}{2k^2\sin(k\eps_1/2)}.
\end{equation*}
The limit with $\eps_1\to0^+$ is obtained by exchanging the two weights.
On the NCM slice they correspond, respectively, to
$\Omega\to-\ii\omega_0$ and $\Omega\to+\ii\omega_0$ at fixed $R$.
This scaled perturbative quantity is distinct from substituting a vanishing
weight into the unscaled density or free energy.  It equals
$2\pi F_{\rm NS}^{\rm rc}(\eps_1/(2\pi),t)$ in the shorthand normalization
used in section~\ref{sec:contour-unref}.

\section{From Chuang's refined contour to the resolved conifold}
\label{app:chuang}

We specialize the general refined contour formula to the resolved conifold and recover
\eqref{eq:I-ref}.  Chuang's proposal
\cite[eq.~(60)]{Chuang:2025aaa} for the non-perturbative refined free energy of a
Calabi--Yau threefold reads
\begin{equation}
 \cF=\sum_{\beta>0,j_L,j_R}N^\beta_{j_Lj_R}\oint_{\cC}\frac{\dd u}{u}\,
 \frac{e^{-(\beta\cdot\boldsymbol t-2\pi\ii n_\beta)u}}{1-e^{-2\pi\ii u}}\,
 \frac{\chi_{j_L}\!\left(e^{\ii u(\eps_1-\eps_2)/2}\right)
       \chi_{j_R}\!\left(e^{\ii u(\eps_1+\eps_2)/2}\right)}
      {4\sin(u\eps_1/2)\sin(-u\eps_2/2)},
 \label{eq:chuang}
\end{equation}
Here Chuang's variables have been written as
$\eps_1=\lambda\bpar$ and $\eps_2=-\lambda/\bpar$, while
$n_\beta=\lfloor\operatorname{Im}(\beta\cdot\boldsymbol t)/(2\pi)\rfloor$
labels the B-field chamber separately for each charge.  For the conifold's
primitive class $\beta_0$ on the thermal slice
$\operatorname{Im}t=\pi$, one has $n_{\beta_0}=0$.
No common zero value is assumed for all curve degrees in a general geometry.
The resolved conifold has
a single rigid $(-1,-1)$ curve carrying one spin-zero hypermultiplet,
$N^{\beta_0}_{0,0}=1$ and all other nonconstant $N^\beta_{j_Lj_R}=0$, so both characters equal one
and \eqref{eq:chuang} collapses to $\oint_\cC\cI\,\dd u$ with $\cI$ as in
\eqref{eq:I-ref}.  The poles are at $u\in\mathbb{Z}_{>0}$,
$u\in(2\pi/\eps_1)\mathbb{Z}_{>0}$ and
$u\in(-2\pi/\eps_2)\mathbb{Z}_{>0}$,
matching the three families of figure \ref{fig:poles}, and the integer residues
reproduce \eqref{eq:ref-GV}.

% ===========================================================================

% ===========================================================================
\section{The endpoint residue, local normalization, and Barnes form}
\label{app:transform}

This appendix supplies the endpoint calculation and normalization comparison
used in subsection~\ref{sec:endpoint}.  The finite-part prescription
\eqref{eq:P-half-finite-part} is compatible with the original
three-derivative identity because $\partial_t^3\cI$ has only a simple pole at
the origin:
\begin{equation}
 \lim_{\rho\to0}\int_{\gamma_0(\rho)}
 \partial_t^3\cI\,\dd u
 =\ii\pi\operatorname{Res}_{u=0}\partial_t^3\cI
 =\partial_t^3P_{1/2}.
 \label{eq:P-half-third-derivative}
\end{equation}
To evaluate the undifferentiated finite part, use
\begin{equation}
 \begin{aligned}
 \frac{1}{1-e^{-2\pi\ii u}}
 &=\frac{1}{2\pi\ii u}+\frac12+\frac{\ii\pi}{6}u+O(u^3),\\[1mm]
 \frac{1}{4\sin(\eps_1u/2)\sin(-\eps_2u/2)}
 &=-\frac{1}{\eps_1\eps_2u^2}
 -\frac1{24}\left(\frac{\eps_1}{\eps_2}
                  +\frac{\eps_2}{\eps_1}\right)+O(u^2),\\[1mm]
 e^{-tu}&=1-tu+\frac{t^2u^2}{2}-\frac{t^3u^3}{6}+O(u^4).
 \end{aligned}
 \label{eq:origin-factor-expansions}
\end{equation}
Including the factor $1/u$ in \eqref{eq:I-ref} and extracting the
coefficient of $u^{-1}$ gives
\begin{equation}
 \begin{aligned}
 P_{1/2}(t;\eps_1,\eps_2)
 ={}&\frac{t^3}{12\eps_1\eps_2}
 -\frac{\ii\pi t^2}{4\eps_1\eps_2}
 -\frac{\pi^2t}{6\eps_1\eps_2}\\
 &+\frac{t-\ii\pi}{48}
 \left(\frac{\eps_1}{\eps_2}+\frac{\eps_2}{\eps_1}\right).
 \end{aligned}
 \label{eq:P-half}
\end{equation}
On the thermal branch,
\begin{equation}
 t=2\pi R\mu+\ii\pi,
 \qquad
 \eps_1\eps_2=-(2\pi R)^2\alpha_+\alpha_- ,
 \label{eq:thermal-branch-data}
\end{equation}
this becomes
\begin{equation}
 \begin{aligned}
 P_{1/2}(t;\eps_1,\eps_2)
 ={}&\frac{(t-\ii\pi)^3+\pi^2(t-\ii\pi)}
          {12\eps_1\eps_2}\\
 &+\frac{t-\ii\pi}{48}
 \left(\frac{\eps_1}{\eps_2}+\frac{\eps_2}{\eps_1}\right).
 \end{aligned}
 \label{eq:P-half-thermal}
\end{equation}

The cubic coefficient can be checked directly from the short-proper-time
region of the NCM kernel.  Since
\begin{equation}
 \sinh\!\left(\frac{\tau}{2R}\right)
 \sinh\!\left(\frac{\alpha_+\tau}{2}\right)
 \sinh\!\left(\frac{\alpha_-\tau}{2}\right)
 =\frac{\alpha_+\alpha_-}{8R}\tau^3+O(\tau^5),
 \label{eq:small-tau-denominator}
\end{equation}
the leading large-$\mu$ contribution at fixed $R$ and $\alpha_\pm$ is
\begin{equation}
 \partial_\mu^3\log Z_{\NCM}^{\rm rot}
 \underset{\mu\to+\infty}{\longrightarrow}
 -\frac{2R}{\alpha_+\alpha_-}
 \int_0^\infty\frac{\sin(\mu\tau)}{\tau}\,\dd\tau
 =-\frac{\pi R}{\alpha_+\alpha_-}.
 \label{eq:large-mu}
\end{equation}
The cubic term in \eqref{eq:P-half-thermal} gives exactly the same result,
\begin{equation}
 \partial_\mu^3
 \left(\frac{(t-\ii\pi)^3}{12\eps_1\eps_2}\right)
 =\frac{(2\pi R)^3}{2\eps_1\eps_2}
 =-\frac{\pi R}{\alpha_+\alpha_-}.
 \label{eq:cubic-origin-check}
\end{equation}
This check fixes only the cubic coefficient; the lower-degree terms follow
from the finite-part prescription itself.

In the unrefined limit $\eps_1=-\eps_2=\lambda$,
\begin{equation}
 P_{1/2}(t;\lambda,-\lambda)
 =-\frac{(t-\ii\pi)^3}{12\lambda^2}
  -\frac{\pi^2(t-\ii\pi)}{12\lambda^2}
  -\frac{t-\ii\pi}{24}.
 \label{eq:unrefined-P-half-thermal}
\end{equation}
The Ho\v{r}ava--Keeler thermodynamic expansion has local part
\begin{equation}
 P_{\rm loc}^{\rm HK}
 =P_2^{\rm HK}(\mu;R,\omega_0,\Lambda)
 -\frac{(t-\ii\pi)^3}{12\lambda^2}
 -\frac{t-\ii\pi}{24},
 \qquad \deg_\mu P_2^{\rm HK}\leq2 .
 \label{eq:HK-local-recalled}
\end{equation}
The cubic and genus-one linear terms therefore agree, while the remaining
linear genus-zero term in \eqref{eq:unrefined-P-half-thermal} is accounted
for by $P_2=P_2^{\rm HK}+\pi^2(t-\ii\pi)/(12\lambda^2)$, as in
appendix~\ref{app:HK-derivation}.

Using the multiple Bernoulli polynomials in the convention of
\cite{Narukawa:2003modular}, the same residue polynomial has the
Barnes representation
\begin{equation}
 P_{1/2}(t;\eps_1,\eps_2)
 =-\frac{\ii\pi}{3!}\,
 B_{3,3}\!\left(
 2\pi\ii-t+\frac{\ii}{2}(\eps_1-\eps_2)
 \;\middle|\;
 2\pi\ii,\ii\eps_1,-\ii\eps_2
 \right).
 \label{eq:barnes}
\end{equation}
Indeed, writing
\begin{equation*}
 \cI(u)=-\frac{e^{Zu}}
 {u\prod_{j=1}^{3}(e^{\omega_ju}-1)},
 \qquad
 \boldsymbol\omega=(2\pi\ii,\ii\eps_1,-\ii\eps_2),
 \quad Z=2\pi\ii-t+\tfrac{\ii}{2}(\eps_1-\eps_2),
\end{equation*}
the generating function
\[
 \frac{u^3e^{Zu}}{\prod_{j=1}^3(e^{\omega_j u}-1)}
 =\sum_{n\geq0}B_{3,n}(Z\mid\boldsymbol\omega)\frac{u^n}{n!}
\]
gives $\operatorname{Res}_{u=0}\cI=-B_{3,3}/3!$.  Equivalently, writing
$W=Z-\tfrac12\sum_j\omega_j=\ii\pi-t$,
\[
 B_{3,3}(Z\mid\boldsymbol\omega)
 =\frac{W^3-\tfrac14 W\sum_j\omega_j^2}{\omega_1\omega_2\omega_3},
\]
which reproduces \eqref{eq:P-half-thermal} directly.

For the comparison with the zero-pole terms of Hattab and Palti
\cite{Hattab:2024chf}, define
\begin{equation}
 \begin{aligned}
 F_0^{\rm Poly}(t)
 &=-\frac16t^3+\frac{\ii\pi}{2}t^2+\frac{\pi^2}{3}t,\\
 F_1^{\rm Poly}(t)
 &=-\frac1{12}t+\frac{\ii\pi}{12}.
 \end{aligned}
 \label{eq:HP-polynomials}
\end{equation}
Then direct substitution into \eqref{eq:P-half} gives
\begin{equation}
 \boxed{
 P_{1/2}(t;\lambda,-\lambda)
 =\frac12\left[
  \frac{F_0^{\rm Poly}(t)}{\lambda^2}+F_1^{\rm Poly}(t)
 \right]
 \equiv P_{\rm HP}^{\rm chamber}(t,\lambda).
 }
 \label{eq:HP-zero-pole}
\end{equation}
The factor $1/2$ has the same two descriptions: the NCM endpoint arcs form a
half-loop around $u=0$, while the full resolved-conifold zero-pole term is
shared by the two flop chambers.

For the normalization comparison, denote the pure-D0/constant-map contribution
introduced in section~\ref{sec:intro} by
\begin{equation}
 \cF_{\rm D0}^{\rm norm}(\eps_1,\eps_2)
 \equiv[\cF_{\rm top}^{\rm full}]_{\beta=0},
 \qquad \partial_t\cF_{\rm D0}^{\rm norm}=0 .
 \label{eq:D0-definition}
\end{equation}
Here $\beta=0$ labels the BPS charge sector, with the local polynomial already
separated.  After choosing the same chamber representative, the unrefined decompositions
are
\begin{equation}
 \begin{aligned}
 \log Z_{\NCM}^{\Omega=0}
 &=\cF_{\rm HP}^{\rm rc,BPS,np}
   +P_{\rm HP}^{\rm chamber}+P_2,\\
 \cF_{\rm top}^{\rm rc,norm}
 &=\cF_{\rm HP}^{\rm rc,BPS,np}
   +P_{\rm HP}^{\rm chamber}+\cF_{\rm D0}^{\rm norm}.
 \end{aligned}
 \label{eq:unref-decomposition}
\end{equation}
Hence
\begin{equation}
 \log Z_{\NCM}^{\Omega=0}-\cF_{\rm top}^{\rm rc,norm}
 =P_2-\cF_{\rm D0}^{\rm norm},
 \label{eq:remaining-normalization-difference}
\end{equation}
which contains only the thermodynamic integration ambiguity and the
modulus-independent pure-D0 normalization.  Neither changes the
nonconstant curve-sector identities proved here.  A specified pure-D0
sector may have additional, modulus-independent Stokes data; these are
not determined by the present comparison.

\section{Non-perturbative residues and Stokes discontinuities}
\label{app:np-res}

At $u_\ell^{(1)}=2\pi\ell/\eps_1$ the factor $\sin(\eps_1u/2)$ has a simple zero
with $\dd\sin(\eps_1u/2)/\dd u=(\eps_1/2)(-1)^\ell$, and the remaining factors
are evaluated at $u_\ell^{(1)}$.  Using
$u_\ell^{(1)}\cdot4\cdot(\eps_1/2)=4\pi\ell$ and
$\sin(-\eps_2u_\ell^{(1)}/2)=\sin(-\pi\ell\eps_2/\eps_1)$,
\begin{equation}
 \Res_{u=u_\ell^{(1)}}\cI
 =\frac{(-1)^\ell e^{-tu_\ell^{(1)}}}
 {4\pi\ell\left(1-e^{-2\pi\ii u_\ell^{(1)}}\right)
  \sin(-\pi\ell\eps_2/\eps_1)},
\end{equation}
which is \eqref{eq:res-np} after multiplying by $2\pi\ii$ and writing
$u_\ell^{(1)}=2\pi\ell/\eps_1$ in the exponentials.  The second family follows by
exchanging $\eps_1\leftrightarrow-\eps_2$.  Using
$\eps_1=\lambda\bpar$ and $-\eps_2=\lambda/\bpar$ gives the weights
$e^{-2\pi\ell t/(\lambda\bpar)}$ with collision factor
$\sin(\pi\ell/\bpar^2)$, and the exchanged family gives
$e^{-2\pi\ell\bpar t/\lambda}$ with $\sin(\pi\ell\bpar^2)$.

It is also useful to absorb the phase in the KK/D0 denominator before
estimating the physical exponential scale.  The same two residues are
\begin{equation}
 \begin{aligned}
 2\pi\ii\Res_{u=u_\ell^{(1)}}\cI
 &=\frac{(-1)^\ell e^{-2\pi\ell(t-\ii\pi)/\eps_1}}
 {4\ell\sin(-\pi\ell\eps_2/\eps_1)
             \sin(2\pi^2\ell/\eps_1)},\\
 2\pi\ii\Res_{u=u_\ell^{(2)}}\cI
 &=\frac{(-1)^\ell e^{2\pi\ell(t-\ii\pi)/\eps_2}}
 {4\ell\sin(-\pi\ell\eps_1/\eps_2)
             \sin(-2\pi^2\ell/\eps_2)}.
 \end{aligned}
 \label{eq:shifted-oscillator-residues}
\end{equation}
On the physical slice $t-\ii\pi\in\mathbb R$ and
$-\eps_2=\overline{\eps_1}$, the second line is the complex conjugate of
the first.  The integer residues are real there as well, since their sine
product is $|\sin(k\eps_1/2)|^2$ and $e^{-kt}=(-1)^ke^{-2\pi R\mu k}$.
The residue decomposition therefore preserves the reality of the NCM
proper-time integral.

For completeness, the same residue sums can be reorganized into three copies of a
single resolved-conifold block.  Define
\begin{equation}
 \mathcal B(t;\eps_1,\eps_2)
 \equiv\sum_{k\geq1}\frac{e^{-kt}}
 {4k\,\sin(k\eps_1/2)\sin(-k\eps_2/2)} .
 \label{eq:B-block}
\end{equation}
Using
\[
 \frac{1}{1-e^{-4\pi^2\ii\ell/\eps_1}}
 =\frac{e^{2\pi^2\ii\ell/\eps_1}}
 {2\ii\sin(2\pi^2\ell/\eps_1)}
\]
in the first line of \eqref{eq:res-np}, and the exchanged identity in the second,
gives
\begin{equation}
 \begin{aligned}
 \oint_{\cC}\cI(u;t,\eps_1,\eps_2)\,\dd u
 ={}&\mathcal B(t;\eps_1,\eps_2)\\
 &+\mathcal B\!\left(t_1^\vee;
       \frac{4\pi^2}{\eps_1},\frac{2\pi\eps_2}{\eps_1}\right)
 +\mathcal B\!\left(t_2^\vee;
       -\frac{4\pi^2}{\eps_2},\frac{2\pi\eps_1}{\eps_2}\right),\\[1mm]
 t_1^\vee={}&\frac{2\pi}{\eps_1}(t-\ii\pi)+\ii\pi,
 \qquad
 t_2^\vee=-\frac{2\pi}{\eps_2}(t-\ii\pi)+\ii\pi .
 \end{aligned}
 \label{eq:three-frames}
\end{equation}
The first block is the integer-pole contribution.  The second and third are the two
oscillator-pole sums after rescaling the period triple
$(2\pi,\eps_1,-\eps_2)$ so that $\eps_1$ or $-\eps_2$, respectively, becomes the
thermal period.  Equation~\eqref{eq:three-frames} is initially valid where the three
series converge separately and elsewhere by analytic continuation; at commensurate
periods the unsplit higher-order residues must be used instead.
Three-period factorizations also underlie the nonperturbative
partition-function proposal of Lockhart and Vafa \cite{Lockhart:2012superconformal}.
\eqref{eq:three-frames} is the nonzero-pole conifold identity in the present
conventions, with its local polynomial kept separate.

% ===========================================================================

\subsection{The double-pole residue and the unrefined limit}
\label{app:unrefined-residues}

For $\eps_1=-\eps_2=\lambda$, let
$u_\ell=2\pi\ell/\lambda$ and
$g(u)=e^{-(t-\ii\pi)u}/[2\ii u\sin(\pi u)]$.
There is no simple-pole term in the Laurent expansion of the squared
oscillator factor itself:
\[
 \frac{1}{[2\sin(\lambda u/2)]^2}
 =\frac{1}{\lambda^2(u-u_\ell)^2}+\frac1{12}
  +O((u-u_\ell)^2).
\]
Consequently $\Res_{u=u_\ell}\cI_{\rm unref}=g'(u_\ell)/\lambda^2$.
Taking the logarithmic derivative of $g$ gives the explicit result
\begin{equation}
 2\pi\ii\Res_{u=u_\ell}\cI_{\rm unref}
 =-\frac{e^{-2\pi\ell(t-\ii\pi)/\lambda}}
          {2\lambda\ell\sin(2\pi^2\ell/\lambda)}
 \left[t-\ii\pi+\frac{\lambda}{2\pi\ell}
                  +\pi\cot(2\pi^2\ell/\lambda)\right].
 \label{eq:unrefined-residue-expanded}
\end{equation}
Differentiating the bracket in \eqref{eq:unref-double-residue} at fixed $t$
produces exactly \eqref{eq:unrefined-residue-expanded}, including the term
from the derivative of the sine denominator.  This independently checks
the sign, the $\lambda$ prefactor, and the argument $t-\ii\pi$ in the
Hattab--Palti expression.

The refined answer approaches this double-pole result only after the two
simple residues are combined.  To see this without subtracting individually
large terms, keep $t$ and $\lambda$ fixed, set
$\eps_1=\lambda\bpar$, $-\eps_2=\lambda/\bpar$, and choose a small
circle $D_\ell$ containing the pair
$2\pi\ell/(\lambda\bpar)$, $2\pi\ell\bpar/\lambda$, but no integer
pole or other oscillator pole.  On this fixed circle the integrand has a
regular limit as $\bpar\to1$, so
\begin{equation}
 \lim_{\bpar\to1}2\pi\ii
 \left(\Res_{u=u_\ell^{(1)}}\cI+
       \Res_{u=u_\ell^{(2)}}\cI\right)
 =\lim_{\bpar\to1}\oint_{D_\ell}\cI\,\dd u
 =2\pi\ii\Res_{u=u_\ell}\cI_{\rm unref}.
 \label{eq:coalescing-residues}
\end{equation}
The apparent divergences of the separated families therefore cancel.
If an integer pole also collides, it must be included inside the same
circle and the resulting triple-pole residue evaluated directly.
This local argument does not require exchanging a singular parameter
limit with either infinite residue sum.

\subsection{Lateral decomposition and the quantum-dilogarithm jump}
\label{app:stokes-derivation}

We now derive the result quoted in subsection~\ref{sec:stokes}.  Parameterize
\begin{equation}
 \eps_1=\lambda\bpar,
 \qquad -\eps_2=\frac{\lambda}{\bpar},
 \qquad \lambda^2=-\eps_1\eps_2,
 \label{eq:lambda-b-param}
\end{equation}
and begin in the noncommensurate reference chamber
\begin{equation}
 \lambda>0,
 \qquad \bpar>0,
 \qquad \operatorname{Re}t>0,
 \qquad \bpar^2\notin\mathbb Q.
 \label{eq:stokes-reference-chamber}
\end{equation}
The displayed residue series are first used where they converge, and
elsewhere with the continuation inherited from the contour.  Irrationality
prevents exact collisions but by itself does not control arbitrarily small
sine denominators.  All image decompositions below are performed with a
common positive origin cutoff before its finite part is taken.
On the lower and upper lateral rays, respectively,
\begin{equation}
 \frac{e^{-tu}}{1-e^{-2\pi\ii u}}
 =\begin{cases}
  \displaystyle\sum_{n\geq0}e^{-(t+2\pi\ii n)u},
  &\operatorname{Im}u<0,\\[3mm]
  \displaystyle-\sum_{n\geq1}e^{-(t-2\pi\ii n)u},
  &\operatorname{Im}u>0.
 \end{cases}
 \label{eq:KK-lateral-expansions}
\end{equation}
The minus sign in the second line is cancelled by the opposite orientation
of the upper component of \eqref{eq:contour}.  Thus
\begin{equation}
 \oint_{\cC}\cI\,\dd u
 =\sum_{n\geq0}\int_{\gamma_-}\omega_n
  +\sum_{n\leq-1}\int_{\gamma_+}\omega_n,
 \label{eq:contour-image-decomposition}
\end{equation}
where
\begin{equation}
 \begin{aligned}
 \gamma_\pm&:0^+\longrightarrow\infty e^{\pm\ii0},\\
 \omega_n(u)
 &\equiv\frac{\dd u}{u}
 \frac{e^{-A_nu/(2\pi)}}
 {4\sin(\lambda\bpar u/2)\sin(\lambda u/2\bpar)},\\
 A_n&\equiv2\pi(t+2\pi\ii n),
 \qquad n\in\mathbb Z.
 \end{aligned}
 \label{eq:image-one-form}
\end{equation}
The image form also identifies the Borel actions directly.  Set
$s=\lambda A_nu/(2\pi)$; then
\begin{equation}
 \omega_n=e^{-s/\lambda}\,\widehat\omega_n(s)\,\dd s,
 \qquad
 \widehat\omega_n(s)
 =\frac{1}{4s\sin(\pi\bpar s/A_n)\sin(\pi s/(\bpar A_n))}.
 \label{eq:image-Borel-kernel}
\end{equation}
Near the origin,
\[
 \widehat\omega_n(s)
 =\frac{A_n^2}{4\pi^2s^3}
  +\frac{\bpar^2+\bpar^{-2}}{24s}+O(s).
\]
Subtracting these genus-zero and genus-one terms leaves an ordinary
Borel kernel for the higher-genus tail and does not change any nonzero-pole
residue.  Its singularities are
$s=\ell A_n/\bpar$ and $s=\ell\bpar A_n$, making the two Stokes directions
explicit.  These image subtractions are only a device for defining lateral
integrals; they do not replace the endpoint prescription for the full
KK/D0-summed kernel in appendix~\ref{app:transform}.

For a fixed image, use the same local prescription on both sides and set
\begin{equation}
 \cF_n^\pm\equiv\int_{\gamma_\pm}\omega_n,
 \qquad
 \operatorname{Disc}_{A_n}\cF\equiv\cF_n^--\cF_n^+.
 \label{eq:stokes-discontinuity-definition}
\end{equation}
Since $\gamma_--\gamma_+$ closes counterclockwise,
\begin{equation}
 \operatorname{Disc}_{A_n}\cF
 =2\pi\ii\sum_{u_*\in\mathcal P_n}
 \operatorname{Res}_{u=u_*}\omega_n,
 \label{eq:disc-as-residues}
\end{equation}
with oscillator poles
\begin{equation}
 u_\ell^{(1)}=\frac{2\pi\ell}{\lambda\bpar},
 \qquad
 u_\ell^{(2)}=\frac{2\pi\ell\bpar}{\lambda},
 \qquad \ell\geq1.
 \label{eq:image-oscillator-poles}
\end{equation}
Their residues are
\begin{equation}
 \begin{aligned}
 2\pi\ii\operatorname{Res}_{u=u_\ell^{(1)}}\omega_n
 &=\frac{\ii(-1)^\ell}{2\ell}
 \frac{e^{-\ell A_n/(\lambda\bpar)}}
      {\sin(\pi\ell/\bpar^2)},\\[1mm]
 2\pi\ii\operatorname{Res}_{u=u_\ell^{(2)}}\omega_n
 &=\frac{\ii(-1)^\ell}{2\ell}
 \frac{e^{-\ell\bpar A_n/\lambda}}
      {\sin(\pi\ell\bpar^2)}.
 \end{aligned}
 \label{eq:image-oscillator-residues}
\end{equation}
Consequently,
\begin{equation}
 \operatorname{Disc}_{A_n}\cF
 =\ii\sum_{\ell\geq1}\frac{(-1)^\ell}{\ell}
 \left[
  \frac{e^{-\ell A_n/(\lambda\bpar)}}
       {2\sin(\pi\ell/\bpar^2)}
  +\frac{e^{-\ell\bpar A_n/\lambda}}
       {2\sin(\pi\ell\bpar^2)}
 \right].
 \label{eq:stokes-residue-series}
\end{equation}

For Faddeev's noncompact quantum dilogarithm \cite{Faddeev:1995modular},
we use the convention of \cite[appendix~A]{Alexandrov:2023refined}:
\begin{equation}
 \Phi_{\bpar}(z)
 =\exp\!\left[
  \int_{\mathbb R+\ii0}\frac{\dd w}{w}
  \frac{e^{-2\ii zw}}
       {4\sinh(\bpar w)\sinh(w/\bpar)}
 \right],
 \label{eq:Faddeev-definition}
\end{equation}
with the logarithm defined by this integral.  Initially take
$\operatorname{Re}\bpar,\operatorname{Re}(\bpar^{-1})>0$ and
$|\operatorname{Im}z|<\tfrac12\operatorname{Re}(\bpar+\bpar^{-1})$;
other values are reached by analytic continuation.  For
$z=-A_n/(2\pi\lambda)$ in a decay chamber, closing the $w$ contour in the
upper half-plane picks the poles $w=\ii\pi\ell/\bpar$ and $w=\ii\pi\ell\bpar$ and gives
\begin{equation}
 \log\Phi_{\bpar}\!\left(-\frac{A_n}{2\pi\lambda}\right)
 =-\ii\sum_{\ell\geq1}\frac{(-1)^\ell}{\ell}
 \left[
  \frac{e^{-\ell A_n/(\lambda\bpar)}}
       {2\sin(\pi\ell/\bpar^2)}
  +\frac{e^{-\ell\bpar A_n/\lambda}}
       {2\sin(\pi\ell\bpar^2)}
 \right].
 \label{eq:Faddeev-residue-series}
\end{equation}
Comparison with \eqref{eq:stokes-residue-series} proves
\eqref{eq:stokes}.  The lower-minus-upper convention fixes the displayed
minus sign; no additional local polynomial is included in this difference.

There is a useful independent check at $\bpar=1$.  Each image now has double
oscillator poles, with smooth numerator
$e^{-A_nu/(2\pi)}/u$.  Its derivative gives
\begin{equation}
 \begin{aligned}
 \left.\operatorname{Disc}_{A_n}\cF\right|_{\bpar=1}
 &=-\frac{\ii}{2\pi}\sum_{\ell\geq1}
   \frac{e^{-\ell A_n/\lambda}}{\ell^2}
   \left(1+\frac{\ell A_n}{\lambda}\right)\\
 &=\left.\frac{\partial}{\partial\lambda}\right|_{A_n}
   \left[\frac{\lambda}{2\pi\ii}
   \Li_2\!\left(e^{-A_n/\lambda}\right)\right]
 =-\log\Phi_1\!\left(-\frac{A_n}{2\pi\lambda}\right).
 \end{aligned}
 \label{eq:unrefined-Stokes-limit}
\end{equation}
This is the unrefined Stokes jump of Hattab--Palti
\cite[eq.~(3.2)]{Hattab:2024chf}.  It differs from
\eqref{eq:unref-double-residue} because an individual image no longer
contains the KK/D0 denominator.  Equation~\eqref{eq:unrefined-Stokes-limit}
also verifies that the singular terms in the two refined families cancel
before their unrefined limit is taken.

The collinear derivation corresponds to
\begin{equation}
 \Omega=\ii\nu,
 \qquad |\nu|<\omega_0,
 \qquad
 \alpha_\pm=\omega_0\mp\nu>0,
 \qquad
 \bpar^2=\frac{\omega_0-\nu}{\omega_0+\nu}>0.
 \label{eq:positive-real-domain}
\end{equation}
For real $\Omega$, choose $\bpar$ continuously from $1$; then
$|\bpar|=1$ and $\operatorname{Re}\bpar>0$.  The two $u$-plane pole
lattices rotate to conjugate rays, while the Borel rays are directed by
$A_n/\bpar$ and $\bpar A_n$.  Continuing the contour gives
\eqref{eq:physical-b-domain} and the combined two-ray transformation
\eqref{eq:combined-two-ray-disc}.  At
commensurate periods the separately written residue sums are replaced by the
combined higher-order residues of the unsplit contour.

% ===========================================================================
\bibliographystyle{unsrt}
\bibliography{references_v15}

\end{document}